# Do Firearm Markets Comply with Firearm Restrictions? How the Massachusetts Assault Weapons Ban Enforcement Notice Changed Registered Firearm Sales

December 2021

Meenakshi Balakrishna and Kenneth C. Wilbur[*]
University of California, San Diego

How well do firearm markets comply with firearm restrictions? The Massachusetts Attorney General issued an Enforcement Notice in 2016 to announce a new interpretation of the key phrase "copies and duplicates" in the state's assault weapons ban. The Enforcement Notice increased assault rifle sales by 1,349 (+560%) within five days, followed by a reduction of 211 (-58%) over the next three weeks. Assault rifle sales were 64-66% lower in 2017 than in comparable earlier periods, suggesting that the Enforcement Notice reduced assault weapon sales but also that many banned weapons continued to be sold.

Acknowledgements and disclaimer:
We gratefully acknowledge the Massachusetts Department of Criminal Justice Information Services for providing data; Arnab Pal for first identifying the data source; Snehanshu Tiwari, Ganesh Chandrasekaran Iyer, Jie Chen, Immanuel Kwok, and Rebecca Wang for superlative research assistance; and John Donohue, Jessica Kim, Kanishka Misra, Yi-Lin Tsai and Robert Sanders for helpful comments and suggestions. Any remaining errors are ours alone.

* Balakrishna: Doctoral Student, mkbalakr@ucsd.edu. Wilbur: Professor of Marketing and Analytics, kennethcwilbur@gmail.com. University of California, San Diego, 9500 Gilman Dr, Box 0553, San Diego, CA 92093.

# I. INTRODUCTION

Western societies use firearm policy to balance the right of self-protection with the risk of enabling deadly force against innocent people. Incidents with four or more gunshot victims have increased rapidly in the U.S., from 270 in 2014 to 611 in 2020 (Gun Violence Archive 2021). Mass shooters that use "assault weapons"—a controversial label that connotes rapid firing, rapid reloading and other combat-related capabilities—wound and kill more people (de Jager et al. 2018, Koper 2020). The U.S. President called for a renewed assault weapon ban in 2021.

Assault weapon restrictions can only have effects if firearm markets comply. Media accounts have indicated widespread non-compliance with past assault weapon restrictions (Atmonavage 2019, Edelman 2015, Haar 2014, King 1991, Mydans 1990). For example, one year after New Jersey made possession of unregistered assault weapons a felony, only 2,000 of an estimated 100,000-300,000 assault weapons had been registered or turned in (King 1991). Two years after New York required registration of existing assault weapons, fewer than 45,000 of an estimated one million assault weapons had been registered (Edelman 2015).[1]

The purpose of the current paper is to estimate the causal effects of a natural experiment in firearm policy on firearm sales. We are particularly interested in what we can learn about firearm market compliance with firearm restrictions, as there is limited direct scientific evidence on the topic. We study an unusual event—the Massachusetts Assault Weapons Ban Enforcement Notice ("EN")—that was announced without prior warning or debate, and apparently prompted a large, immediate, short-lived spike in rifle sales. The EN defined key terms in Massachusetts' legal definition of banned assault weapons, thereby indicating that the ban applied to many popular weapons.

---

[1] These case studies are suggestive but lack complete measurements and control groups; and they describe noncompliance with restrictions on existing firearms, rather than restrictions on new firearm acquisitions.

Massachusetts requires all firearm dealers to register every new firearm sale with the state. Figure 1 graphs the population of legally registered firearm sales, by type of firearm, for a 6-week period centered on the date of the Enforcement Notice. Two features stand out. First, Rifle sales rose more than 20-fold on the date of the EN, without comparable increases in Handgun or Shotgun sales. Second, the series of Rifle sales after the peak appears to be only slightly lower than before the Enforcement Notice. Of course, the graph does not show what sales would have been observed in the absence of the EN; simple pre/post comparisons can be muddled by multiple factors such as seasonality.

We estimate causal effects of the EN on firearm sales. We focus on legal firearm sales as an outcome variable because the express purpose of the assault weapon ban and EN are to restrict legal acquisitions of new assault weapons. We are particularly interested in sales of restricted assault weapons so we can gauge market compliance with the EN. We employed research assistants to classify market-leading rifles according to the state's legal definition of assault weapons. This enables us to distinguish four types of rifle sales: assault weapons, non-assault weapons, ambiguous, and unclassified. We then used a semi-supervised machine learning procedure to identify and classify additional sales of the same weapons.

Next, we estimate auto-regressive models to predict counterfactual sales, i.e., what sales would have occurred in the absence of the EN, based on lagged sales, time effects and firearm license issuances by date. Firearm licenses are good instruments for short-term firearm sales: licenses are required to legally purchase or possess a firearm in Massachusetts, and application processing times ensured that license issuances in the first few weeks after the EN could not have been caused by the EN itself. We compare post-EN firearm sales to the model's predictions to estimate immediate and short-run causal effects. We find that assault rifle sales rose sharply in the first few days after the EN, an increase almost entirely due to repeat firearm purchasers. Assault rifle sales declined substantially in the following three weeks.

Finally, we present a descriptive analysis of longer-term trends in firearm sales after the EN. If the assault weapon classifications are perfect, and if the market complied fully with the EN, then assault rifle sales should have been zero in 2017. However, we find that 2017 assault rifle sales were 64-66% lower than comparable earlier periods. The evidence shows a level shift downward, but no downward trend within 2017. Hence, the longer-term analysis suggests partial effectiveness of the EN: it seems to have reduced assault weapon sales substantially, but it also shows that many banned assault weapons continued to sell.

Overall, the results indicate the degree to which the Massachusetts firearm market complied with the Assault Weapon Ban Enforcement Notice, confirming that new assault weapon restrictions would be likely to reduce assault weapon sales. However, we view the reduction as an upper bound for the true effect, as there is no way to know how many illegal transactions may have occurred without recording in the data. Further, the observed noncompliance suggests that new regulations may require enforcement to achieve full compliance.

We proceed by describing the relevant literature, the empirical context of the natural experiment and the details of the policy change. We then describe the data and firearm classification procedure, methods and findings, and longer-term data descriptives. We conclude with limitations and policy implications.

## II. PREVIOUS LITERATURE

Broadly, we contribute to a literature that estimates causal effects of firearm policy changes on behavior. Raissian (2016) found that the 1996 Gun Control Act, which prohibited defendants found guilty of domestic violence misdemeanors from possessing firearms, reduced gun-related homicides among female intimate partners and children. Luca, Malhotra, and Poliquin (2017) found that handgun waiting periods reduced gun homicides in the states that pass them. Donohue, Aneja and Weber (2019) found that right-to-carry concealed handgun laws increased overall violent crime.

Narrowly, we study how a change in assault weapon policy affects firearm markets and policy compliance. A highly relevant series of studies of assault weapon markets was funded by Congress as part of its 1994 ten-year ban on new assault weapons and large-capacity magazines (Roth and Koper 1997, 1999; Koper and Roth 2001, 2002; Koper, Woods and Roth 2004; see also Koper 2013). A surge in assault weapon production immediately preceded the ban, and a surge in assault weapon prices immediately followed it, but prices returned to pre-ban levels within a few years. The assault weapon ban reduced the number of assault weapons used in crimes, mostly due to a reduction in assault pistols, but the reduction was mostly offset by increasing use of other semiautomatic weapons. The ban on new large-capacity magazines was more consequential in crime reduction, though its effect did not take hold for several years due to the large existing stocks at the time of the ban. This work implies that the EN-induced surge in assault weapon sales documented below may be unlikely to have led to new firearm crimes.[2]

---

[2] Koper, Woods and Roth (2004) noted several aspects of their analysis that relate to the current paper. They pointed out that their time-series analyses were "largely descriptive, so causal inferences must be made cautiously." They also explained, "our studies of the [assault weapon] ban have shown that the reaction of manufacturers, dealers, and consumers to gun control policies can have substantial effects on demand and supply for affected weapons both before and after a law's implementation. It is important to study these factors because they affect the timing and form of a law's impact on the availability of weapons to criminals and, by extension, the law's impact on gun violence." Such considerations were "largely absent from prior research" on firearm policies.

Another closely related study by Koper (2014) linked new firearm acquisitions to guns used in crimes. The analysis connected 72,000 legal firearm sales in Maryland to a national database of crime guns recovered by police, identifying more than 1,800 firearms present in both data sets. Semiautomatic weapons were disproportionately likely to be recovered by police as crime guns. A state regulation of the used firearm market did not appear to change the probability of a gun's recovery, possibly due to unclear enforcement of the statute.

Studdert et al. (2017) examined the reaction of legal firearm sales in California to two highly publicized mass shootings in Newtown and San Bernardino. They found the two shootings increased handgun sales by 53% and 41%, respectively. The increases were larger among first-time handgun purchasers. The San Bernardino shooting increased handgun acquisitions by 85% in San Bernardino, well above the 35% increase in the rest of the state.

Kim (2021) examined how a series of expansive firearm usage policies—permitless concealed carry, shall-issue concealed carry licensing, and "stand your ground" self-defense rules—affected gun sales and public health outcomes. She found that shall-issue licensing shifted demand from long guns to lower-priced handguns, and that permitless concealed carry coincided with increased firearm assaults and firearm suicides.

The topic of firearm market compliance with regulations has received limited attention, perhaps because noncompliance is often difficult to measure. To our best knowledge, the most closely related paper is Castillo-Carniglia et al. (2018), who studied how three states' adoption of comprehensive background check policies changed firearm background checks. Background checks increased in Delaware but not in Colorado or Washington. Two related explanations for the non-increases are non-compliance and non-enforcement, as the new policies led to anti-regulation rallies, protests and statements by law enforcement authorities. Relatedly, Kravitz-Wirtz et al. (2021) surveyed firearm owners in California, finding appreciable fractions of firearm owners self-reported purchasing weapons without background checks in violation of the state's background check laws.

We supplement the indirect and self-reported evidence of non-compliance with direct, observational evidence. This theme relates to "forensic economics," which Broulik (2020) describes as "economics informing any enforcement stage in any field of law," and Zitzewitz (2012) describes by saying "the economist sizes the extent of an activity about which there had been only anecdotal evidence and provides insight into where it is more prevalent and why." Our approach to forensic evidence is based on detecting behavior that is "hidden in plain sight," in the sense that the data we analyze were provided directly to the state by firearm retailers, some of whom participated in transactions that contravened the EN.[3]

To the best of our knowledge, the policy event we study has not received any prior scientific attention.[4] We provide direct causal evidence of how firearm policy changed firearm sales. We think this evidence might help policy makers, in Massachusetts and elsewhere, understand the consequences of the Enforcement Notice. Second, we believe this paper provides some of the first indication of the rapidity of firearm market participants' response to restrictive firearm policies, even when they are announced without prior warning. Third, we provide long-run descriptive evidence that the Enforcement Notice reduced sales of semiautomatic assault weapons, but that many banned weapons continued to be sold.

---

[3] Christie and Schultz (1994) and Christie, Harris and Schultz (1994) used a similar approach to detect collusion by stock market makers. Fisman and Miguel (2007) examined why some United Nations officials in Manhattan left parking tickets unpaid and how officials responded to a change in parking ticket enforcement policies.
[4] The closest analogue might be other analyses of sudden policy changes, such as Cunningham and Shah (2018), who found that sudden decriminalization of sex work in Rhode Island increased the prostitution market while simultaneously reducing rape reports and gonorrhea incidence.

All comprehensive reviews of the firearm literature agree that more research is needed, including research on how policy influences the market for firearms. As IOM and NRC (2013) explained, "Basic information about gun possession, acquisition, and storage is lacking." They further noted that "fundamental questions about the effectiveness of interventions—both social and legal—remain unanswered." Similarly, Smart et al. (2020, Recommendation 9) suggested that, "to improve understanding of outcomes of critical concern to many in gun policy debates, the U.S. government and private research sponsors should support research examining the effects of gun laws on a wider set of outcomes, including … the gun industry."

## III. EMPIRICAL CONTEXT: THE NATURAL EXPERIMENT

*A. Background: Assault Weapon Restrictions*

Mass shootings have sometimes precipitated changes in firearm policy (Luca, Poliquin and Malhotra 2019), some of which have reduced mass shootings (Donohue and Boulouta 2019; Lemieux, Prenzler and Bricknell 2013). Assault weapons have been restricted or banned in many countries, including the United States (1994-2004),[5] Australia (1996-present), New Zealand (2019-present), Canada (2020-present), and most of Europe.

Seven U.S. states ban new assault weapon sales. 67% of U.S. adults surveyed said that they favor banning assault-style weapons (Pew Research 2017), and the U.S. President called for a renewed assault weapons ban in 2021. Large retail chains Walmart and Dick's Sporting Goods have voluntarily stopped selling assault weapons. At the same time, many other states are loosening firearm restrictions (Kim 2021) and numerous ongoing court cases challenge the constitutionality of specific firearm restrictions.[6]

---

[5] The U.S. federal government also banned new machine guns in 1986, importations of semiautomatic rifles in 1989 and 1998, sales of new semiautomatic assault weapons from 1994 until 2004, and bump stocks in 2018.

When the U.S. federal government banned "semiautomatic assault weapons" in 1994, it defined them as the following categories of firearms:[7]

- 9 specific firearm products (e.g., Colt AR-15, AK-47, etc.), including "copies and duplicates" of those specific products.
- Semiautomatic rifles that accept detachable magazines and offer at least two of five features: folding/telescoping stock, pistol grip, bayonet mount, flash suppressor, and grenade launcher.
- Semiautomatic pistols and semiautomatic shotguns with at least two features among similar feature sets.

The definition reflected a political compromise (Lenett 1995). Several of the specific firearm products, most notably the Colt AR-15, were already off patent and readily extensible when the ban was passed in 1994. After the statute was enacted, firearm manufacturers quickly introduced minor variations on the specific firearm products named in the ban, thereby apparently skirting the "copies and duplicates" restriction. Many variants were based on the Colt AR-15, the semiautomatic version of the fully automatic M16 machine gun. Derivatives of the AR-15 are called "AR-15-style," "AR-style," or "Modern Sporting Rifles." An estimated 1-2 million AR-style rifles were manufactured in 2016[8] with 8.5-15 million in circulation.[9] Colt retired its consumer-market AR-15 in 2019, stating that "the market for modern sporting rifles has experienced significant excess manufacturing capacity."[10]

---

[6] An unusual sequence of events in Boulder, Colorado, in March 2021 illustrated possible links between regulations, shootings, and policy reactions. The city's assault weapons ban was overturned on March 12 by a court ruling that it contravened state law. On March 23, a mass shooter killed ten people with an AR-556. The Colorado state legislature passed two firearm-related bills in April 2021, and introduced several more in May 2021.

[7] H.R. 3355-203, Title XI, sec. 110102. https://www.congress.gov/bill/103rd-congress/house-bill/3355/text

[8] https://web.archive.org/web/20190821183313/https://www.washingtonpost.com/news/business/wp/2018/02/23/u-s-gun-manufactures-have-produced-150-million-guns-since-1986/

[9] https://www.mcclatchydc.com/news/nation-world/national/article201882739.html

*B. The Massachusetts Assault Weapon Ban Enforcement Notice*

Massachusetts banned assault weapons in 1998 using the same definition of assault weapon as the 1994 federal assault weapons ban.[11] Unlike the federal law, the Massachusetts ban did not expire, and therefore has operated continuously since its enactment. First-time offenders face fines of $1,000-$10,000 and imprisonment of 1-10 years.[12]

On the morning of July 20, 2016, the Massachusetts Office of the Attorney General (AGO) issued a public Enforcement Notice (EN) regarding prohibited assault weapons.[13] The EN was announced in a press conference and a *Boston Herald* editorial and generated substantial immediate publicity in the state. The AGO also notified the state's firearm retailers of the change in registered letters it mailed in the afternoon of July 18, 2016.

The EN explained what the AGO considered to be "copies or duplicates" of the specific firearms banned under the state's legal definition of assault weapons. The phrase "copies or duplicates" was not defined in the 1994 federal statute or the 1998 state law and had not been interpreted by the courts. The EN introduced two new tests that indicate whether a firearm is a "copy or duplicate" of a banned weapon: a *Similarity Test* indicating whether its internal functional components are substantially similar to a banned firearm, and an *Interchangeability Test* indicating whether its receiver is interchangeable with the receiver of a banned firearm. The 1998 state law explicitly banned the Colt AR-15, so the Enforcement Notice thereby indicated that the Massachusetts assault weapons ban covered many popular AR-style rifles, among other weapon types.

---

[10] https://www.nytimes.com/2019/09/19/business/colt-ar-15.html
[11] https://www.nraila.org/articles/20160722/massachusetts-attorney-general-unilaterally-bans-thousands-of-previously-legal-guns
[12] https://malegislature.gov/Laws/GeneralLaws/PartI/TitleXX/Chapter140/Section131M
[13] https://www.mass.gov/files/documents/2018/11/13/assault-weapons-enforcement-notice.pdf

Contemporaneous AGO communications further indicated that the AGO (i) would immediately apply the Enforcement Notice to new firearm sales; (ii) would not enforce the EN retroactively to firearms sold prior to July 20, 2016, and (iii) did not plan to enforce the EN on relevant weapon transactions that had been initiated prior to July 20, 2016, but not yet been completed.[14] Because the EN applied only to new weapon sales that had not yet been started, it did not affect owners' existing assault weapons and would not prevent legal transfers or sales of existing assault weapons.

The AGO did not provide a list of firearm products that were newly banned under the Enforcement Notice. AGO communications accompanying the EN said "The Attorney General expects voluntary compliance from gun dealers and manufacturers with respect to prohibited weapons," and that, "By issuing the notice, the Attorney General hopes and expects that non-compliant gun dealers will come into voluntary compliance with the law, to minimize the need for criminal or civil enforcement."[15]

The EN announcement was widely seen as a surprise. It was not preceded by public comment or debate. Contemporaneous news accounts indicate that firearm owners were given "little to no notice," that firearm purchasers "flocked to" firearm retailers on the EN announcement date, and that an online petition protesting the EN received over 13,000 signatures overnight (Sacharczyk 2016, Lipovich 2016).

---

[14] https://www.mass.gov/guides/frequently-asked-questions-about-the-assault-weapons-ban-enforcement-notice
[15] https://www.mass.gov/guides/frequently-asked-questions-about-the-assault-weapons-ban-enforcement-notice

The EN came 39 days after a mass shooting in which 53 people died in Orlando on June 13, 2016. Highly publicized mass shootings are known to stimulate immediate legal firearm acquisitions (Liu and Wiebe 2019, Studdert et al. 2017). Several other shootings and newsworthy events proximate to the EN may have affected firearm sales: 11 police officers were shot in Dallas on July 7, followed by smaller police shootings in California on July 26 and 29; the California governor signed several gun control bills and vetoed several others in early July 2016; the Republican National Convention took place in Cleveland from July 18-21; there was a mass shooting in Munich that killed 9 people on July 23; and the Democratic National Convention took place in Philadelphia from July 25-28.

The Enforcement Notice was followed by public debate about whether the AGO had exceeded its legal authority. The EN has since been affirmed by multiple judicial decisions with additional litigation ongoing.[16]

## C. Identification Strategy, Hypotheses, Instrumental Variables

The EN and related publicity apparently caused a dramatic and immediate reaction in firearm sales, as shown in Figure 1. Given the unanticipated nature of the Enforcement Notice, we assume its timing to be an exogenous shock to the Massachusetts firearm market. We therefore treat it as a "natural experiment" or "quasi-experiment." We use its timing to identify the causal effect of a change in assault weapon policy on firearm sales in the immediate and short run.

The natural experiment is unusual for two reasons. First, a change in the interpretation of a firearm ban occurred without an accompanying change in the statute itself. We are not aware of

---

any other occasion on which the legal definition of a banned firearm changed during the time the ban was continually in effect. Therefore, we can separate the causal effects of the policy change from any related effects of other factors such as legislative debate, news coverage, or firearm industry marketing policies.

Second, although it is straightforward to predict that a broader interpretation of banned assault weapons should reduce assault weapon sales, it is less obvious to predict how large the effect might be. The size of the effect may indicate the extent to which firearm market participants comply with firearm restrictions. It is also unclear how sales of non-banned firearms would react. For example, if a broader interpretation of banned assault weapons leads purchasers to substitute other firearms for assault weapons, then sales of other firearms might rise.

Licenses are required to legally purchase or possess a firearm in Massachusetts. They cost $100, are available to both residents and nonresidents, and must be renewed every six years in order to legally acquire or maintain possession of a firearm. License applications take up to 42 days to be processed.[17] State government personnel told us that 35-40 days normally pass between license application and issuance. There is no fast-track or expedited licensing procedure.[18] We use firearm license data to improve predictions of counterfactual firearm sales for several weeks after the EN. The justification for the license instrument is that, because the EN announcement was a surprise, license applications before the EN could not have been driven by the EN. Therefore, license issuances for several weeks after the EN must be unrelated to EN timing.

---

[17] https://goal.org/ltc-apply/
[18] Massachusetts passed a law phasing out Class B licenses in 2014, but its effect was minor, as Class A licenses and Firearm Identification Cards account for 96% of all licenses issued from 2006-2017, and anywhere from 95-97% of all licenses in every year from 2011 to 2017.

We limit the time span of the econometric exercise to 25 days after the EN, during the period when firearm licenses should offer a strong instrument. We do not extend the causal-effects analysis beyond that limited threshold because no strong instrument or control group is available for the increase in firearm sales during the fall 2016 election season. However, we do report longer-term data descriptives which suggest that the EN was partially but imperfectly effective in reducing assault weapon sales on a longer horizon.

## IV. DATA

*A. Provenance*

Massachusetts state law requires firearm dealers to record all firearm sale transaction data—including date, purchaser, firearm type, Make and Model—in a state-owned database. We obtained anonymous firearm sales and license data by filing a Freedom of Information Act request with the Massachusetts Department of Criminal Justice Information Services.

We focus on the population of 954,015 legally registered dealer sales that occurred between Jan. 1, 2006, and April 1, 2018. 755,688 sales occurred prior to the EN. Each record indicates the transaction date, Firearm Make, Firearm Model, Firearm Type (Handgun, Rifle or Shotgun),[19] Dealer Name, Dealer Shop City, an anonymous purchaser identifier and purchaser gender.

Handguns accounted for 61.3% of all sales, followed by Rifles (26.8%) and Shotguns (11.8%). We do not analyze gender as purchasers were overwhelmingly male at 93%. We do not observe transaction prices.

---

[19] The state also tracks Machine Gun sales, but they account for only 0.03% of the transactions and are subject to distinct federal restrictions on transfer and ownership, so we exclude them from the analysis.

*B. Classifying Rifle Sales by Type*

We seek to distinguish the effect of the EN on banned assault weapons from its effect on other types of firearms. We focus on Rifles because Handgun and Shotgun sales did not respond substantially to the EN. We encountered several challenges:

1. The data do not directly indicate which firearms are assault weapons.

2. There were 44,710 unique Rifle Make/Model combinations listed in the data. This is because firearm sales are highly dispersed across Makes and Models, and also because the Make and Model data contain ambiguities, spelling errors and omissions.

3. Many records provide insufficient information to reach conclusive classifications. The most prominent example is the Ruger Mini-14, the fourth-highest-selling Rifle Make/Model combination in 2016 with 462 transactions. Ruger offered many variants of the Mini-14, including the "Mini-14 Ranch" variants which are semi-automatic, but do not accept detachable magazines or meet the two-feature test for assault weapons. On the other hand, the Ruger Mini-14 Tactical 5888 clearly meets the legal definition of assault weapons, as it is semiautomatic, accepts detachable magazines, and includes three relevant features: folding stock, pistol grip and flash suppressor.[20] Some Ruger Mini-14 sales records indicate which variant was sold, but many records do not indicate the variant. Therefore, the Make/Model combination "Ruger Mini-14" is inherently ambiguous because accurate classification requires the oft-missing variant data.

We took two steps to derive clearer signals from the data. First, we employed three research assistants (RAs) to classify rifles with high sales in 2016 as assault weapons or not.

---

[20] https://web.archive.org/web/20201112021945/https://ruger.com/products/mini14TacticalRifle/specSheets/5888.html , accessed August 2021.

Second, we extended those classifications to matching transaction records using a semi-supervised machine learning procedure. We describe each step in turn.

*Step 1. Research assistant coding*

We provided the legal definition of semiautomatic assault rifles to the RAs and asked them to independently develop a full understanding of its component parts (e.g., "semiautomatic," "detachable magazine," "folding or telescoping stock," etc.). We discussed the definitions with them to ensure that they had successfully followed the instructions. Then we provided a list of the 98 highest-selling unique Rifle Make/Model combinations in 2016, which collectively accounted for 14,367 Rifle sales in 2016, or 50.0% of all 2016 records with non-unique Make/Model combinations.

We asked the RAs to search online for manufacturer specifications, product listing pages and product reviews for each Make/Model combination. Then, based on their best understanding, indicate whether each Rifle Make/Model combination was a semiautomatic assault weapon according to Massachusetts' legal definition; not a semiautomatic assault weapon; or indeterminate due to missing information. The RAs worked independently without a time limit. They recorded and provided online sources of information for each Rifle so we could evaluate the information they used in reaching their judgments. Our manual review of the RAs' classifications and information sources made us confident that they took the task seriously and performed it competently. As a result we view unanimous RA agreements as an authoritative and reasonably stringent classification of assault weapons.

Next, we developed and applied the following summary classifications to Rifle Make/Model combinations:

*AWR-True*: All three RAs indicated that Rifle features conclusively meet Massachusetts' legal definition of assault weapon.

*AWR-False*: All three RAs indicated that Rifle features conclusively do not meet Massachusetts' legal definition of assault weapon.

*AWR-Ambig.*: The three RAs did not all agree on a conclusive indication.

*AWR-Unclassified*: Any weapon the RAs did not classify.

Table 1 provides the list of 98 rifles classified and their 2016 sales. 40 of 98 Rifle Make/Model combinations were classified AWR-True, accounting for 6,170 total sales in 2016. 27 Rifle Make/Model combinations were classified AWR-False, accounting for 2,856 sales records in 2016. 31 Rifle Make/Model combinations were classified AWR-Ambig, accounting for 5,341 sales records in 2016.

We did not alter the state's data prior to creating the RA classifications because we did not want to bias the RAs' judgments. As a result, the Rifle Make/Model combinations included several obvious duplicates, e.g. the Smith & Wesson "M P 15," "M P-15," "M P15," "M P 15-22," etc.[21]

*Step 2. Semi-supervised Machine Learning to Extend RA Classifications to Mistaken Records*
We extended the RA classifications to sales records of unambiguously matching Rifle Make/Model combinations. First, we reduced the number of unique Make values. Make field data contained numerous abbreviations, typos and misspellings. A common example is "Smith & Wesson" commonly abbreviated as "SW" or "S & W". Another common example is "Henry," "Henry Arms," "Henry Repeating," "Henry Repeating Arms," "Henry Repeating Arms Co.", and "Henry Repeating Arms Company" which obviously all refer to the same firearm manufacturer.

---

[21] Table 1 also includes the Ruger AR-556, which firearm retailers often entered into the data base as a Rifle, even though legally it is a handgun due to its barrel length.

We manually assigned all frequent Make values to a smaller set of corrected Makes and then double-checked the assignments.

Next, we consolidated duplicate Model entries. Model field errors were more nuanced. Many manufacturers use subtle codes for rifle models, variants and products, as can be seen in Table 1. We consolidated transaction records by performing the following steps for each corrected Make value:

1. We used the R package *fuzzyjoin* to suggest matches of unclassified Model entries to RA-classified weapons. We analyzed all suggestions of matched records that differed by at most one character.

2. We manually reviewed all matched records to check for false positives. We often double-checked both records against the manufacturer's product pages to ensure correctness. We only extended RA classifications to unambiguously correctly matched records.

3. We manually reviewed all unmatched Model records with 5 or more sales transactions in 2016 to check for any false negatives, i.e., unmatched Models within the same Make. Again, we often checked manufacturer product information to ensure correctness. We only extended RA classifications to unambiguously correct matches.

We manually double-checked steps 2 and 3 to ensure correctness of all classified records. This process extended the RA classifications to an additional 4,258 Rifle sales records, an increase of 30% over the 14,367 sales records of the 98 highest-selling Make/Model combinations in 2016. Overall, 25% of all 2016 Rifle sales transactions are classified as AWR-True, 12% as AWR-False, 18% as AWR-Ambig., and 46% as AWR-Unclassified.

In conclusion, we believe that the AWR-True and AWR-False labels are highly reliable and easily interpretable. We believe the AWR-Ambig data are reliable but less easy to interpret, as the distinction results mainly from missing product and variant information. Finally, the AWR-Unclassified label is easily interpreted as missing data on infrequently-selling weapons, which if they had been classified, would include observations of all three other rifle types.

## C. Trends in Firearm Sales and Licenses

Figure 2 shows daily firearms sales, by date and by firearm type, for a three-month period from June 1 until August 30, 2016. It shows that the Rifle sales spike on the date of the EN announcement (July 20) was almost entirely due to AWR-True and AWR-Unclassified weapons. AWR-Ambig. sales increased but by a relatively smaller proportion and AWR-False weapons showed virtually no increase.

Figure 2 also shows that AWR-True and AWR-Ambig. rifle sales also increased substantially on June 14, 2016, immediately after the Orlando mass shooting, and remained elevated for several weeks afterward. However, the immediate reaction after the Orlando shooting was less pronounced than the EN reaction. Once again, AWR-False and AWR-Ambig. rifle sales did not change nearly as much.

Figure 3 graphs monthly firearm sales by type over the 12-year data set. Annual sales grew more than four-fold from 2007 until 2017, from 27,207 to 111,616. The average annual growth rate was 21% over that period, including a 7% decline in 2014 and a 13% decline in 2017. Handgun sales grew the most and the fastest. Handgun sales increased by 442% over ten years, from 16,612 in 2007 to 73,493 in 2017. The Massachusetts population grew 5% over the same ten-year period.[22]

---

Figure 3 also shows large increases in AWR-True and AWR-Ambig. rifle sales during the Fall 2012 national election, the December 2012 mass shooting in Newtown, CT, and the subsequent renewed legislative debate about gun control. It also shows that, after the EN in July 2016, AWR-True sales dipped substantially, but never fell all the way to zero. AWR-True sales were mostly flat in 2017 before turning upward in early 2018, but did not recover to their May 2016 level during our data set. We explore these patterns further in section VI.

We also observe data on all Massachusetts firearm license issuances, classified according to whether they are new licenses or license renewals.[23] Figure 4 shows license issuances by new/renewal status and year. A license remains valid for six years, a pattern which can be seen clearly in the license renewals time series. From 2006-2011, there were 97,658 new licenses and 207,784 license renewals issued. From 2012-2017, there were 209,993 new licenses and 260,936 license renewals issued. Thus, the number of legally permitted firearm owners increased by about half, from about 4.6% to 6.9% of the state population, during the latter six-year period.

Figure 5 shows that daily license issuances did not change discontinuously around the time of the EN. They reached a local maximum about one week after the EN, a spike that likely corresponded to new applications filed in the wake of the mass shooting in Orlando on June 13, 2016.

*D. Data Validation*

Braga and Hureau (2015) studied related Massachusetts firearm data as part of their study of sources of crime guns recovered by the Boston Police Department. They documented that dealers were unable to locate sales records for 2.2% of trace requests, suggesting imperfections in recordkeeping. And as documented above, we found numerous data entry errors and ambiguities. Therefore, we sought to evaluate the completeness of the Massachusetts firearm sales transaction data using an external source.

---

[23] Resident Class A and Firearms Identification Card licenses account for 96% of all licenses issued. Both license types enable legal firearm purchases and possession.

We compared the Massachusetts records to state/year National Instant Criminal Background Check System (NICS) data reported by the U.S. Federal Bureau of Investigation. NICS data count background checks conducted by federally licensed firearm retailers prior to a firearm purchase, with separate counts for Handguns and Long Guns. Comparing annual Massachusetts sales records to annual FBI NICS data between 2006 and 2017, we find a correlation of 0.97 for Handguns and 0.96 for Long Guns. Therefore, it appears that the Massachusetts data correspond very highly, though not perfectly, to the external FBI data source.[24] This suggests that Massachusetts sales records may be reasonably accurate.

## V. METHODS

### A. Model and Estimation

We predict counterfactual firearm sales—i.e., what sales would have occurred in the absence of the EN—in order to estimate immediate and short-run causal effects of the EN on firearm sales. We estimate regression models using the pre-EN sales data, from January 1, 2006, until July 19, 2016. Causal effects are then calculated as the difference between observed sales and counterfactual predictions.[25]

---

[24] We expect imperfect correspondence because background checks and firearm sales differ. Would-be firearm buyers may fail background checks; a buyer may decide against purchasing a weapon during or after a background check delay; background checks and associated firearm sales may occur in contiguous time periods.

[25] Our empirical approach is conceptually similar to the "interrupted time series" strategy that Gonzalez-Navarro (2013) used to study Lojack introduction on car thefts, except that it uses a single geographic unit rather than a panel of geographies, and there is no available analogue to Gonzalez-Navarro's exposure variable (i.e., the number of cars treated with Lojack that could have been stolen).

The regression models daily firearm sales, for each type of firearm, as a function of variables that could not have reacted to the EN: lagged sales before the EN, time fixed effects and trends, and license issuances, which depended on pre-EN license applications. We specify the following regression model:

$$y_{jt} = \sum_{\tau=1}^{T_y} \alpha_{j\tau} y_{j,t-\tau} + x_t \beta_j + \sum_{\tau=1}^{T_z} \gamma_{j\tau} z_{N,t-\tau} + \sum_{\tau=1}^{T_z} \delta_{j\tau} z_{R,t-\tau} + \varepsilon_{jt} \tag{1}$$

$y_{jt}$ is the log of the number of firearm sales of type $j$ on date $t$;[26] $x_t$ includes a set of time fixed effects described below; $z_{Nt}$ and $z_{Rt}$ are the log of the number of new licenses and license renewals issued on date $t$; and $\varepsilon_{jt}$ is an error term. We expect license data to help predict firearm sales because license issuances and renewals may prompt consumers to consider a new firearm acquisition, or they may predate planned firearm purchases since a currently valid license is needed to buy a gun.

The time effects include weekday fixed effects, a holiday fixed effect, week-of-year fixed effects, year fixed effects and linear and quadratic time trends. The set of time effects chosen minimized prediction errors in a 10-fold cross-validation exercise, relative to alternate specifications. We then chose to include $T_y = 28$ lags of sales by successively adding lags to the model until the first time an additional lag increased the mean absolute prediction error in the validation folds. Then, while including 28 lags of sales, we successively added lags of license variables, finding that $T_z = 10$ lags of license issuances minimized mean absolute prediction error in the validation folds. The results were relatively insensitive to choices of $T_y$ and $T_z$. We experimented with dropping outlying sales values, and found that excluding outliers more than

---

[26] We added 0.1 to firearm sales and license variables before taking logs to avoid taking the log of zero.

five standard deviations above the mean marginally improved the model's predictive performance.[27]

*B. Model Fit, Parameter Estimates, Retrodiction Exercise*

We estimate the model for each firearm type using ordinary least squares. R-square statistics were 0.81 for AWR-True sales, 0.67 for AWR-Ambig., 0.64 for AWR-False, 0.74 for AWR-Unclassified, 0.81 for Handgun, and 0.74 for Shotgun.

The Appendix provides the parameter estimates for each of the 6 regressions. Firearm sales are generally highest on Saturdays and lowest on Sundays and Mondays. New license issuances are strong predictors of firearm sales of all six types, but license renewals significantly predict sales of only three firearm types (AWR-True, Handgun, Shotgun). The year effects, time trend and autoregressive parameter estimates show disparate patterns across the firearm types.

We are primarily interested in the model's predictive performance around the time of the Enforcement Notice. We therefore conducted a retrodiction exercise to compare hold-out predictions to observed sales. For this exercise, we estimated the model without using the final 10 days of data before the EN (July 10—19, 2016). Figure 6 compares the model's sales predictions for July 10-19, 2016, to actual sales observed during the 10-day holdout period.

Although this 10-day window occurred just before the EN, it also occurred 28-37 days after the highly publicized Orlando mass shooting on June 13, and shortly after the Dallas shooting on July 7. Some of the purchases during the 10-day holdout period may have related to those anomalous events. Therefore, this is a challenging test of the model's predictive performance.

---

[27] We also experimented with a Poisson model but that did not improve the model's predictive performance.

The model underpredicted cumulative firearm sales during this 10-day period by 13.8%. Mean daily prediction errors were –9% for AWR-True, -20% for AWR-Ambig, -25% for AWR-False, -8% for AWR-Unclassified, –15% for Handguns and –13% for Shotguns. Daily prediction errors were greatest on July 10 and 11, at 45% and 51%; their range in the following eight days spanned –10% on July 16 to +22% on July 18.

Overall, we conclude that the model is able to predict firearm sales reasonably well immediately before the EN, suggesting that it likely would produce reasonable counterfactual predictions after the EN. However, its mild underestimation of firearm sales immediately before the EN may indicate that post-EN firearm sales predictions are conservative.

## VI. CAUSAL EVIDENCE

We compare observed firearm sales, by type, to the model's predictions of what counterfactual sales would have been in the absence of the EN. We distinguish between an *immediate* effect occurring on the EN announcement date and the four days following, and a *short-run* effect that occurred from 5-25 days after the EN.[28]

Figure 7 graphs actual and predicted sales for the immediate effect. In the five days after the EN, consumers purchased 1,349 (+560%) more *AWR-True* Rifles than predicted, and 1,001 (+431%) more AWR-Unclassified Rifles than predicted. The large majority of the surplus sales occurred on the date of the EN announcement.

The EN had smaller immediate impacts on sales of other firearm types. AWR-Ambig rifle sales exceeded cumulative predictions by 178 (+261%), AWR-False rifle sales increased by

---

[28] We focus on a 25-day window out of concern that post-EN firearm license applications might affect the license issuances instrument. Similar conclusions are reached if we focus on the broader 5-35-day window instead.

37 (+87%) over cumulative predictions, Handgun sales increased by 289 (+28%) in the five-day period after the EN, and Shotgun sales increased by 106 (+83%) after the EN.

How much of the sales spike represents "last-chance" purchasing among firearms enthusiasts or collectors vs. "commercially-motivated" purchases by professional firearms traders who anticipated appreciation on the used firearms market? The 5-day sales spike primarily consists of regular retail buyers: single-transaction purchasers bought 59.6% of Rifles sold during the spike, and two-transaction purchasers bought an additional 22.1%. However, the other end of the distribution may show some commercial motivations as well: 30 buyers bought between 5-15 Rifles each, and a single buyer purchased 27 Rifles, collectively accounting for 8.3% of all Rifle purchases during the spike. The 5-day period immediately before the spike offers a point of reference, when single-transaction purchasers accounted for 86% of all Rifle sales, and buyers with 5 or more purchases accounted for just 3.3% of Rifle purchases.[29]

Figure 8 graphs the short run effects from 5-25 days after the EN, by type of firearm. Observations and model predictions show some correspondence across dates. Cumulative observed AWR-True sales were 211 below the predicted level (-58%). AWR-Unclassified sales were 231 below cumulative predictions (-25%).

Cumulative observed short-run AWR-False sales were 43 (+21%) more than predicted. Cumulative observed short-run AWR-Ambig sales were 127 (+40%) greater than predictions. Handgun sales were 169 (-4%) less than the model predicted. Cumulative observed short-run Shotgun sales were 13 (+2%) more than the cumulative prediction.

---

[29] We also checked whether patterns of retailer sales differed immediately before vs. during the spike. We found that the top 10 retailers accounted for 34.7% of Rifle sales during the 5-day spike, slightly less than the top 10 sellers' 39.1% collective share in the 5-day period preceding the spike.

In summary, the EN caused large immediate increases in AWR-True and AWR-Unclassified Rifle sales, followed by short-run decreases of 58% and 25%, respectively. If the short-run reduction in AWR-True Rifle sales was exact and permanent, it would have taken 6.4 weeks for the negative short-run effect of the EN to counterbalance the positive immediate impact of the EN. If the 25% reduction in short-run AWR-Unclassified Rifle sales was exact and permanent, it would have taken 4.3 weeks for the negative short-run effect of the EN to counteract the positive immediate impact of the EN.

## VII. LONGER-TERM DATA DESCRIPTIVES

There is no control group or instrument available to isolate longer-term causal effects of the EN from other factors (e.g., the 2016 election, mass shootings, etc.). Therefore, we do not attempt to estimate long-run causal effects of the EN, but we present long-run data descriptives that indicate market compliance with the EN.

### A. Firearm Sales in 2017

Figure 9 graphs weekly firearm sales, by type and week of the calendar year, with separate lines for each year from 2014 until 2017. The figures are on the log scale to prevent the post-EN sales spike from distorting the y-axis. The annual overlays show limited seasonality in firearm purchases.

The spikes in AWR-True and AWR-Ambiguous Rifle sales in mid-2016 remain prominent. AWR-True sales fell during the final few months of 2016, a pattern that does not appear for AWR-False or AWR-Ambig. Rifle sales.

In the first half of 2017, AWR-True Rifle sales hovered near the bottom end of the range of 2014-2016 levels for nearly the entire year. Sales in the second half of 2017 were similar to late 2016 in many weeks, and both were typically below comparable 2014 and 2015 levels.

The graph shows a downward shift in the level of AWR-True Rifle sales with fewer large spikes after the EN, but no clear downward time trend within 2017. For example, the change in AWR-True Rifle sales between 2015 and 2017 was -66%. The change in AWR-True Rifle sales in the first half of 2016 vs. the first half of 2017 was similar at -64%. The other five firearm types showed sales patterns that either were similar between 2017 and previous years, or higher in 2017 than in previous years.

The AWR-True Rifle sales decrease of 64-66% between 2017 and comparable earlier periods suggests that the EN partially succeeded in decreasing assault rifle sales in the long run. However, if the AWR-True rifle classification is correct, and if the market complied fully with the EN, then the decrease would be 100%. The substantial gap between a 64-66% observed drop and a hypothetical 100% drop suggests that many illegal firearm sales occurred in 2017.

We have thought about why there may have been noncompliance with the Enforcement Notice. We emphasize that the following is speculation given the absence of process evidence. One possibility is civil disobedience: perhaps many firearm sellers and purchasers knew about the EN and deliberately disregarded it. Another is ignorance or uncertainty: perhaps some market participants either did not know about the EN or did not accept it as legitimate. Third, as the AGO made clear, the AGO relied on dealers' voluntary compliance with the law. Such compliance may have been difficult or unprofitable. Finally, we have found no evidence that assault weapons ban enforcement efforts increased concurrently with the EN. In fact, the AGO publicly requested voluntary compliance, and its communications implied that some dealers were already known to be out of compliance. It may be that enforcement is required to gain a higher level of compliance in the marketplace.

*B. Sales to Newly-observed Purchasers*

How much of the 2016 sales spike was caused by new gun purchasers as opposed to repeat purchasers or collectors? The data contain anonymous purchaser identifiers, so we can partially address this question. We identified the first week each anonymous purchaser identifier appeared in the data, and then counted purchases by that identifier in its first observed week as sales to "newly-observed purchasers." Note that some of the newly-observed purchasers may have purchased weapons in other states or prior to 2006, so this group mixes first-time gun buyers with an unknown proportion of previous gun owners.

Figure 10 graphs weekly sales to newly-observed purchasers, by firearm type and week of year, with separate lines for each year from 2014 until 2017. Newly-observed purchasers accounted for just 2.5% of the EN-related sales spike in AWR-True Rifles, and 7.7% of the EN-related sales spike in AWR-Unclassified Rifles. Thus we conclude that the sales spike occurred on the intensive margin, i.e., existing firearm owners purchasing additional weapons rather then new purchasers buying weapons for the first time.

## VIII. DISCUSSION

We offer rare causal evidence of how firearm policy changed firearm sales. We employed research assistants and a semi-supervised machine learning algorithm to classify rifles as assault weapons. We estimated a model of firearm sales up until the point of the Enforcement Notice, and then measured causal effects of the EN by comparing the model's predictions to observed sales data in the 25 days after the announcement. We found that the Enforcement Notice caused large, immediate increases of sales of AWR-True and AWR-Unclassified Rifles, followed by smaller short-run decreases. The EN did not cause similarly large changes to AWR-False, Shotgun or Handgun sales. AWR-True Rifle sales were about 64-66% lower in 2017 than in comparable earlier periods, suggesting that the EN reduced legal sales of semiautomatic assault rifles, and also that many banned assault weapons continued to be sold.

*A. Limitations*

It is relevant to consider that retailers entered the data into the state database. We do not know of any auditing or verification that is performed by the state. Two data weaknesses seem especially likely. First, transactions may have been backdated to the date of the Enforcement Notice. Deceptive transaction dates might lead us to misinterpret assault weapon sales timing.

A second possibility is that retailers may have declined to enter illegal transactions into the state database. If that occurred, then the effects we reported above would bound the true effects. The possibility of unobserved illegal activities is likely to affect most similar studies of noncompliance.

Other limitations also bear mentioning. The data come exclusively from Massachusetts, a state with unusually strict firearm policy and low firearm ownership, and therefore may not represent other jurisdictions. The data have limited ability to explain the mechanisms underlying the main causal effects. For example, we do not observe prices, EN-related publicity (e.g., emails from firearm retailers to consumers), consumer search or store traffic. Finally, we have not connected the legal purchase data to other possibly related outcomes, such as the used guns market, the illegal firearm market, gun trace data, suicide, homicide or aggravated assault.

*B. Implications for Firearm Policy*

Several results might interest policy makers. First, firearm sales apparently reacted extremely rapidly to firearm policy changes. It may well be that the firearm industry amplifies policy-related news in order to stimulate immediate sales. Policy makers may wish to anticipate market reactions and design their public communication strategies accordingly.

Second, we have showed that the EN was partially effective in reducing assault weapon sales. This effect occurred without any accompanying revision of state law. The result may be relevant to assault weapons ban enforcement and interpretation in other jurisdictions.

Third, feature-based definitions of banned weapons have been criticized as "porous" (Donohue 2012), and have led weapon manufacturers to skirt legal bans by modifying firearm features, such as using replaceable magazines in place of detachable magazines (White 2019). If policy seeks to restrict weapons whose capabilities facilitate mass shootings—such as rapid firing and rapid reloading—it may be advisable to consider restricting weapon capabilities, rather than weapon features. This suggestion is informed speculation; it does not follow directly from the empirical analysis.

Finally, this research has implications for firearm data collection and provision. Massachusetts could audit and monitor its firearm sales data. It could require firearm dealers to enter weapon characteristics to facilitate automated detection of assault weapons. It could publish privacy-compliant firearm sales data to the public. More and better scientific analyses will become possible if more jurisdictions collect and disseminate privacy-compliant firearm sales data.

*C. Conclusion*

In summary, this paper seeks to contribute to the growing body of empirical knowledge about firearm regulation by classifying assault rifles, by analyzing how the EN changed firearm sales, and by providing the first direct, observational measurement of firearm market compliance with firearm restrictions. We are hopeful that the body of empirical firearm research will continue to grow. We need more and better firearm research if we hope to have evidence-based firearm policy.

Disclosures:

Authors have no funding or conflicts of interest to report. We undertook this research with the goal of providing empirical facts to help inform policy makers. As IOM and NRC (2013) put it, "In the absence of research, [firearm] policy makers will be left to debate controversial policies without scientifically sound evidence about their potential effects." We will post final data and estimation code online.

# Table 1. Rifle Classifications

| AWR-True | | | AWR-False | | | AWR-Ambig | | |
|---|---|---|---|---|---|---|---|---|
| *Corrected Make* | *Unique Model Entry* | *2016 Sales* | *Corrected Make* | *Unique Model Entry* | *2016 Sales* | *Corrected Make* | *Unique Model Entry* | *2016 Sales* |
| Ruger | AR-556 | 672 | Remington | 700 | 426 | Ruger | 10-22 | 1636 |
| Windham Weaponry | WW-15 | 476 | Marlin | 60 | 235 | Ruger | MINI-14 | 462 |
| Smith & Wesson | M P 15 SPORT II | 373 | Remington | 783 | 172 | Ruger | AMERICAN | 395 |
| Smith & Wesson | M P 15-22 | 341 | Winchester | 94 | 164 | Springfield Armory | M1A | 236 |
| Smith & Wesson | M P-15 | 273 | Remington | 597 | 147 | Springfield Armory | ARMORY M1A | 233 |
| Ruger | AR556 | 250 | Savage Arms | ANSCHUTZ AXIS | 140 | Ruger | MINI-14 | 229 |
| Smith & Wesson | M P15 SPORT II | 249 | Henry Repeating Arms | H001 | 110 | Mossberg | MVP | 177 |
| Smith & Wesson | M P15-22 | 228 | Savage Arms | ANSCHUTZ AXIS XP | 104 | Beretta | CX4 STORM | 173 |
| Smith & Wesson | M P 15 | 199 | Ruger | 10 22 TD | 102 | Hi Point | POINT 995 | 162 |
| Smith & Wesson | M P15 | 175 | Marlin | 336 | 86 | Mossberg | PATRIOT | 123 |
| Anderson Mfg | AM-15 | 174 | Marlin | XT-22 | 85 | Hi Point | 995 | 120 |
| Mossberg | MMR | 163 | Henry Repeating Arms | GOLDEN BOY | 84 | Iwi | X95 | 94 |
| Smith & Wesson | M P 15-22 SPORT | 161 | Marlin | 336W | 82 | Hi Point | 995TS | 92 |
| Bushmaster | XM15-E2S | 152 | Ruger | 10 22 TAKEDOWN | 78 | Iwi | TAVOR X95 | 82 |
| Iwi | TAVOR | 137 | Henry Repeating Arms | H001 | 75 | Hi Point | POINT 4095 | 80 |
| Aero Precision | X15 | 128 | Ruger | 10 22-RB CARBINE | 75 | Savage Arms | ANSCHUTZ 93R17 | 74 |
| Ruger | PRECISION RIFLE | 128 | Winchester | 70 | 73 | Norinco | SKS | 73 |
| Windham Weaponry | WW15 | 124 | Savage Arms | ANSCHUTZ RASCAL | 72 | Keltec | SUB2000 | 72 |
| Stag Arms | STAG-15 | 119 | Henry Repeating Arms | BIG BOY | 70 | Hi Point | FIREARMS 995 | 72 |
| Smith & Wesson | M P SPORT II | 110 | Russia | M91 30 | 66 | Savage Arms | ANSCHUTZ 11 | 72 |
| Smith & Wesson | MP15-22 | 107 | Savage Arms | ANSCHUTZ MARK II | 63 | Hi Point | 4595 | 71 |
| Smith & Wesson | MP15 | 105 | Savage Arms | ANSCHUTZ A17 | 63 | Ruger | K10 22-TD | 69 |
| Smith & Wesson | 15-22 | 100 | Ruger | 10 22RB | 58 | Smith & Wesson | M P 15 22 | 67 |
| Ruger | PRECISION | 97 | Savage Arms | ANSCHUTZ 10 | 58 | Kriss | VECTOR | 67 |
| Anderson Mfg | AM-15 | 94 | Henry Repeating Arms | H004 | 58 | Ruger | RANCH RIFLE | 65 |
| Anderson Mfg | MFG AM-15 | 89 | Russian State Factories | CENTURY A M91 30 | 56 | Keltec | SUB 2000 | 64 |
| Smith & Wesson | M P15-22 SPORT | 88 | Remington | 742 | 54 | Century Arms | ARMS RAS47 | 62 |
| Mossberg | 715T | 86 | | | | Mossberg | MVP PATROL | 61 |
| Sig Sauer | M400 | 80 | | | | Hi Point | 995TS | 54 |
| Dpms | A-15 | 78 | | | | Hi Point | 4595 | 52 |
| Century Arms | C39V2 | 68 | | | | Smith & Wesson | M P 15 SPORT 2 | 52 |
| Smith & Wesson | M P 15 SPORT | 66 | | | | | | |
| Spikes Tactical | ST15 | 64 | | | | | | |
| Colt | M4 CARBINE | 63 | | | | | | |
| Windham Weaponry | WW15 | 63 | | | | | | |
| Anderson Mfg | AM15 | 62 | | | | | | |
| Colt | LE6920 | 61 | | | | | | |
| Smith & Wesson | MP15 SPORT 11 | 60 | | | | | | |
| Smith & Wesson | M P-15 SPORT II | 55 | | | | | | |
| Sig Sauer | 516 | 52 | | | | | | |

**Figure 1. Rifle sales spike on EN announcement (2016-07-20)**

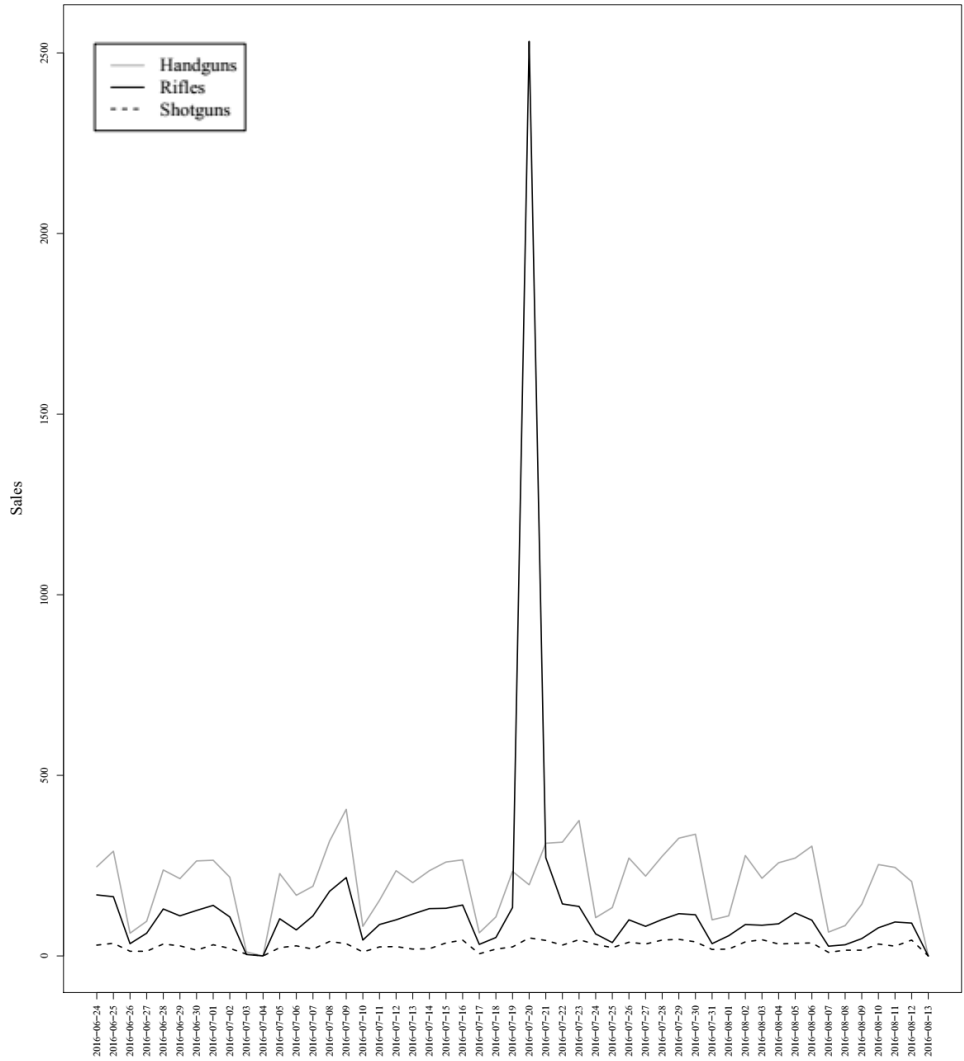

**Figure 2. Firearm Sales by Date and Type, Jun-Aug 2016**

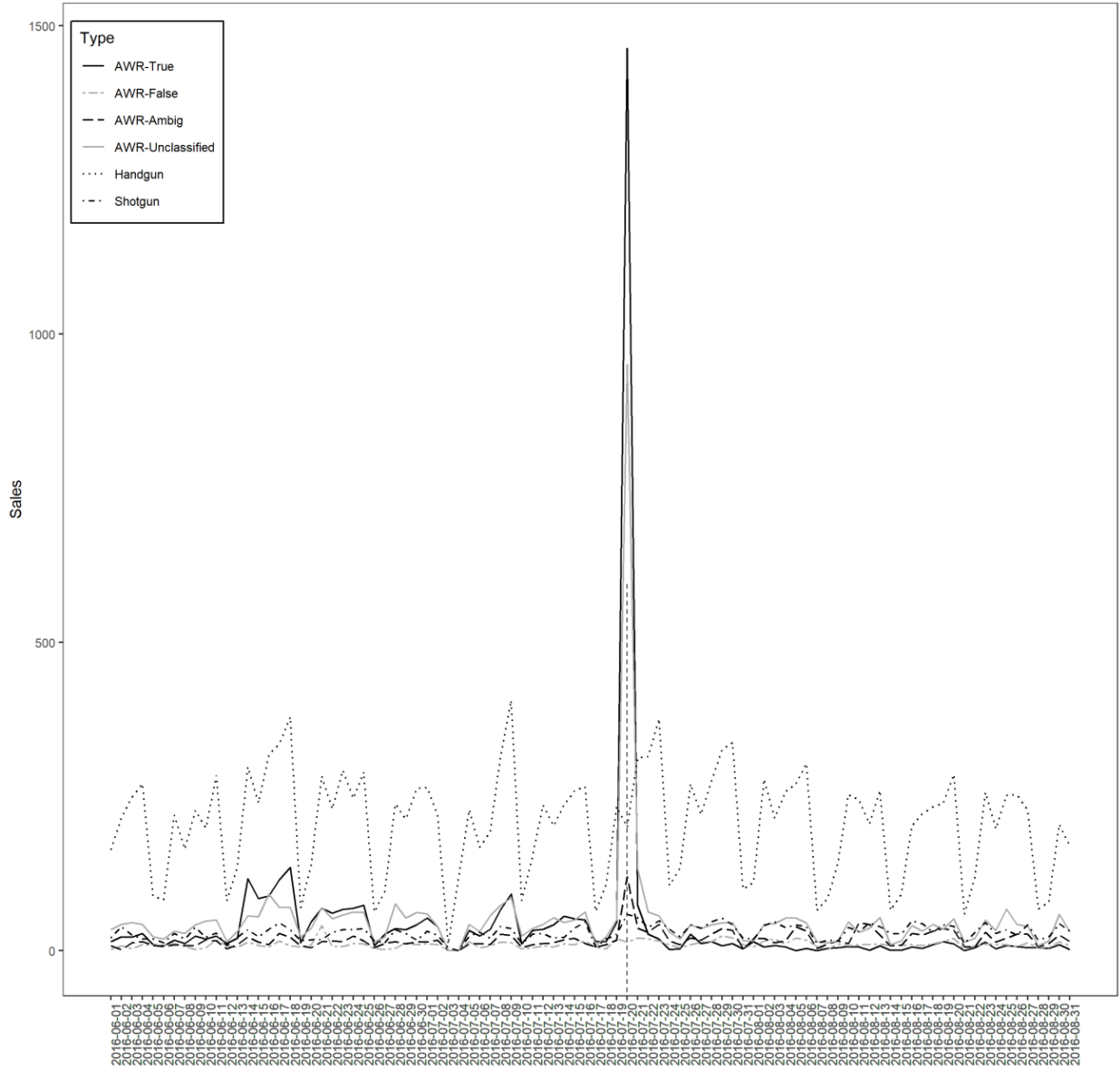

**Figure 3. Firearm Sales by Year/Month and Type**

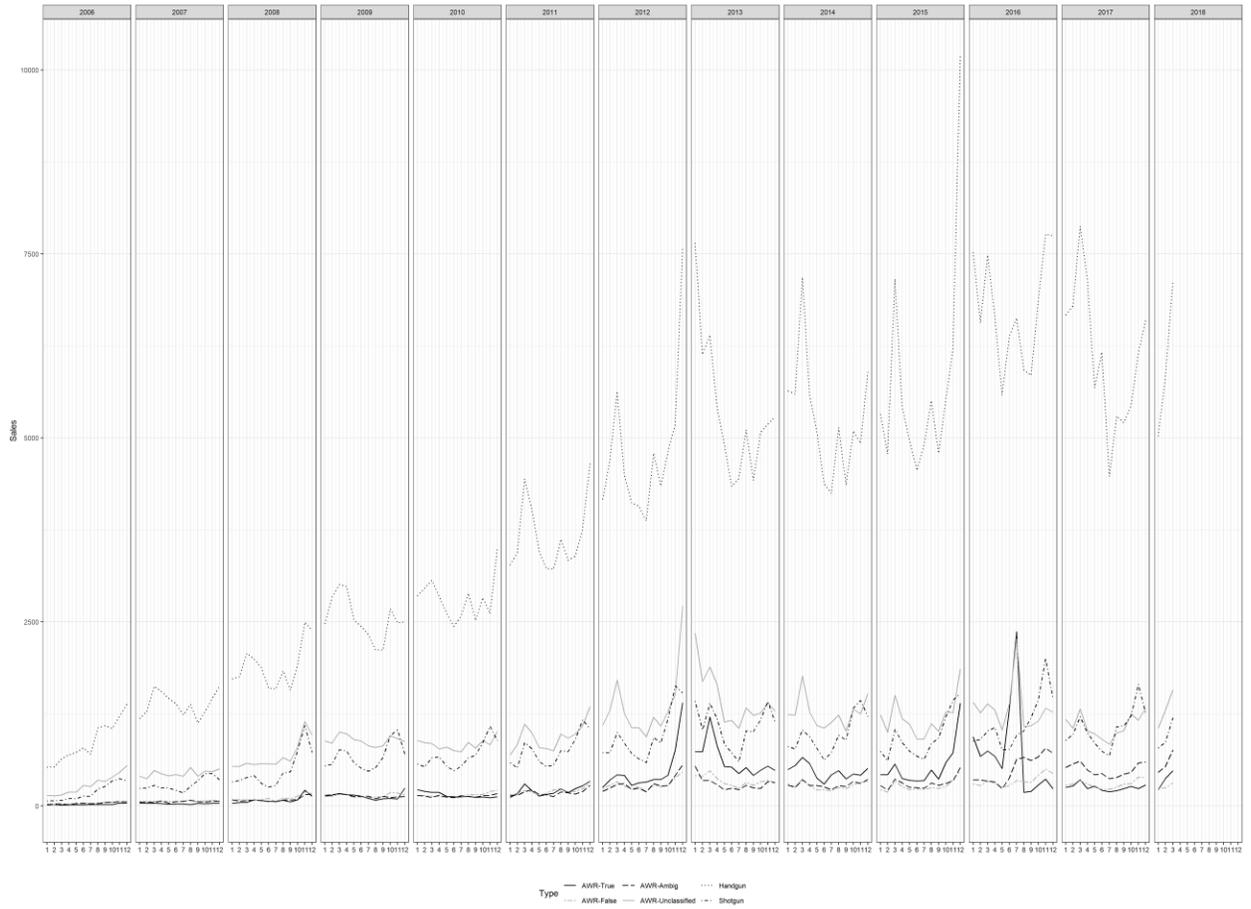

**Figure 4. Annual License Issuances by Type**

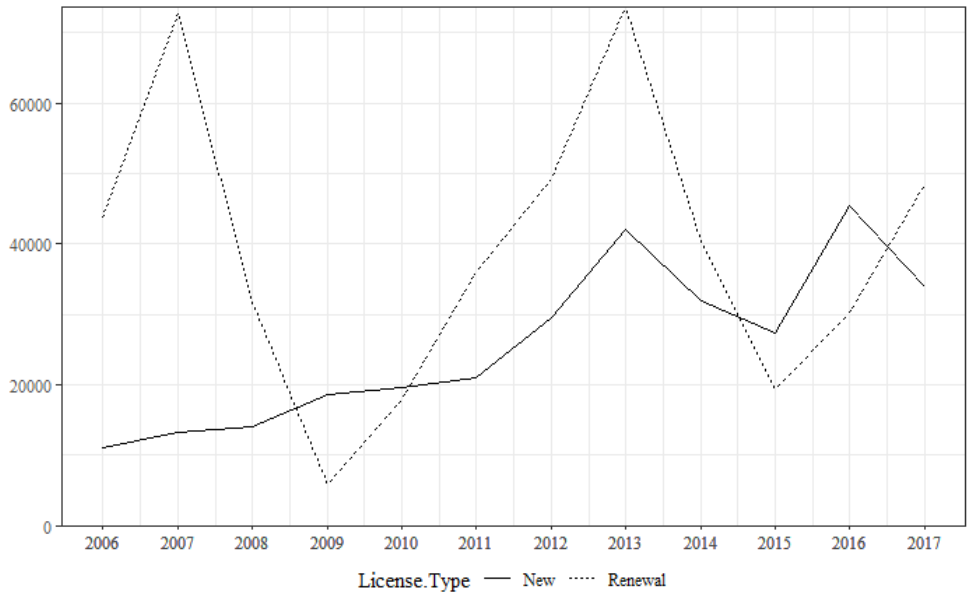

**Figure 5. Daily License Issuances Around the EN Date**

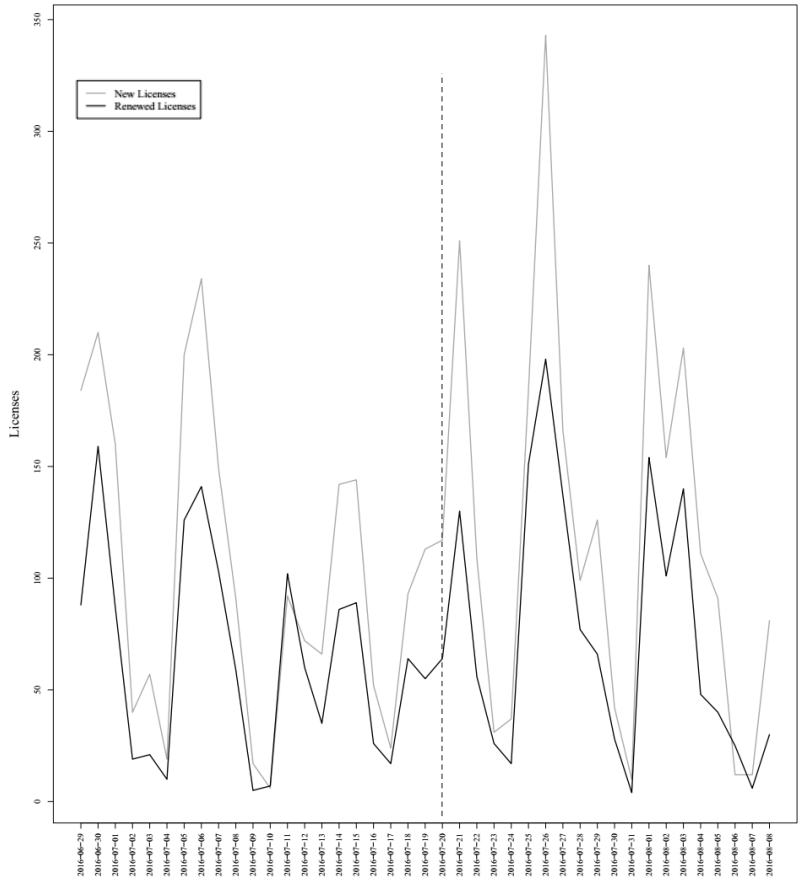

# Figure 6. Pre-EN Retrodictions

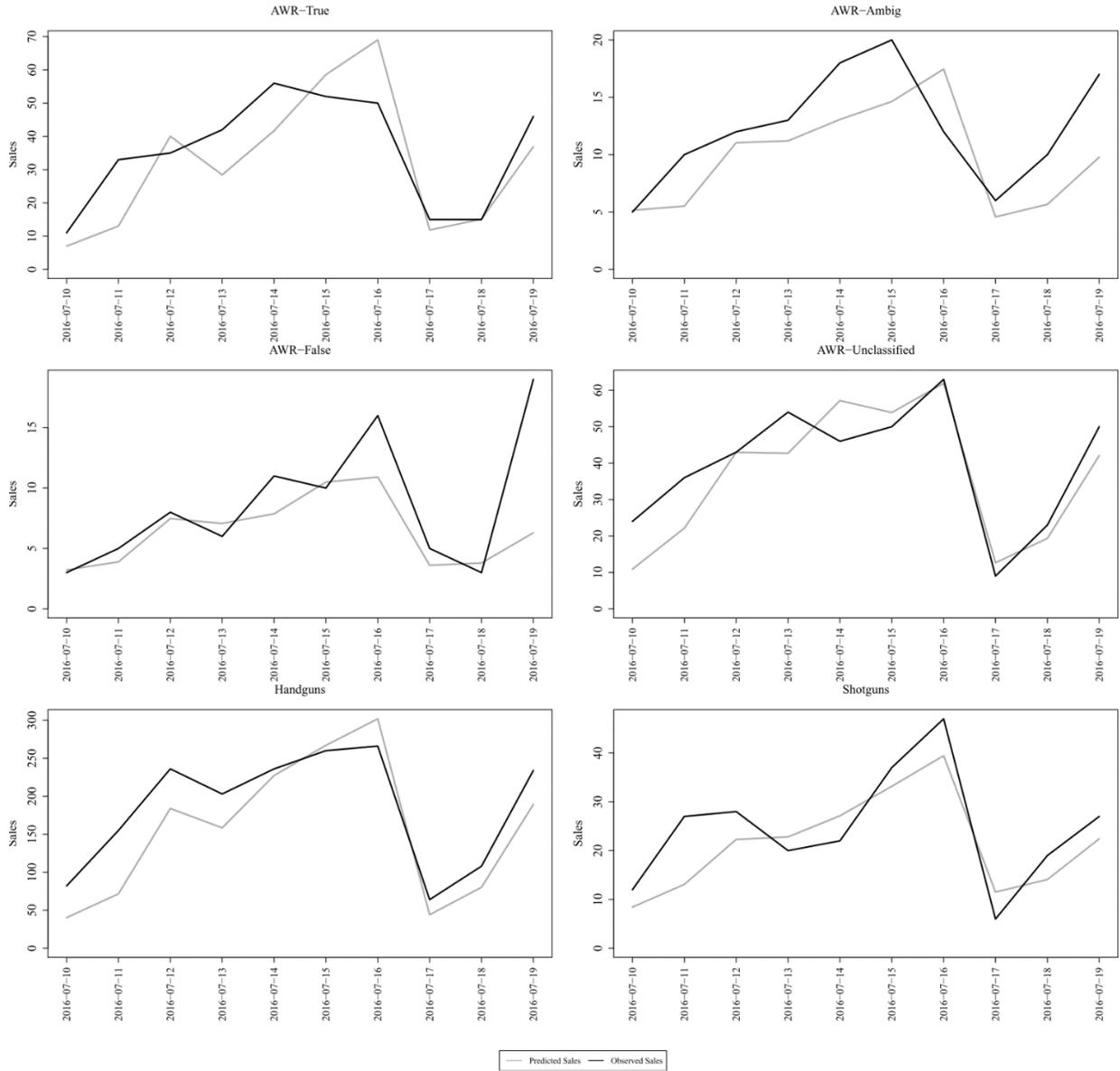

**Figure 7. Immediate effect of EN on Firearm Sales, by Type**

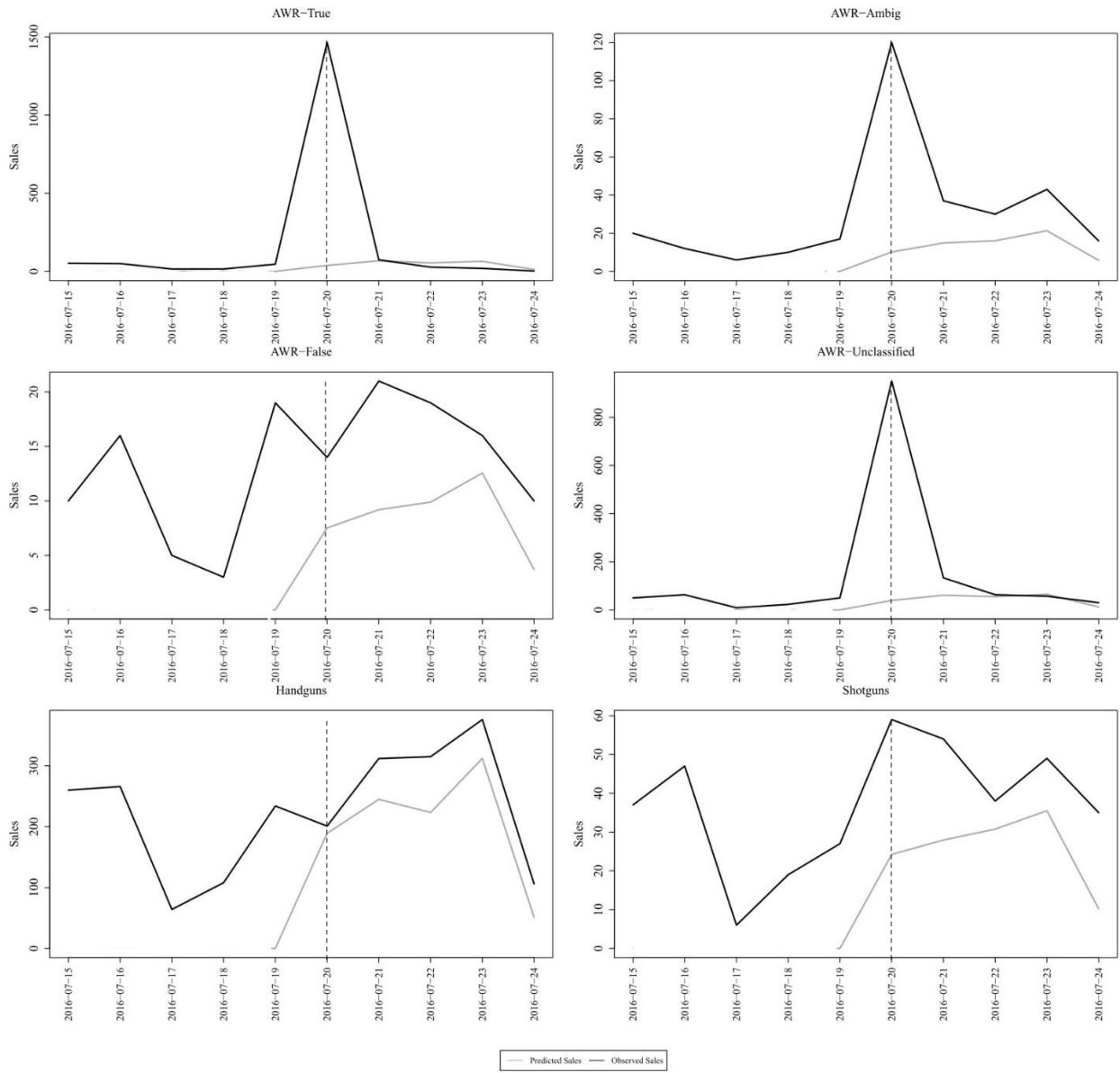

**Figure 8. Short-run effects of EN on Firearm Sales, by Type**

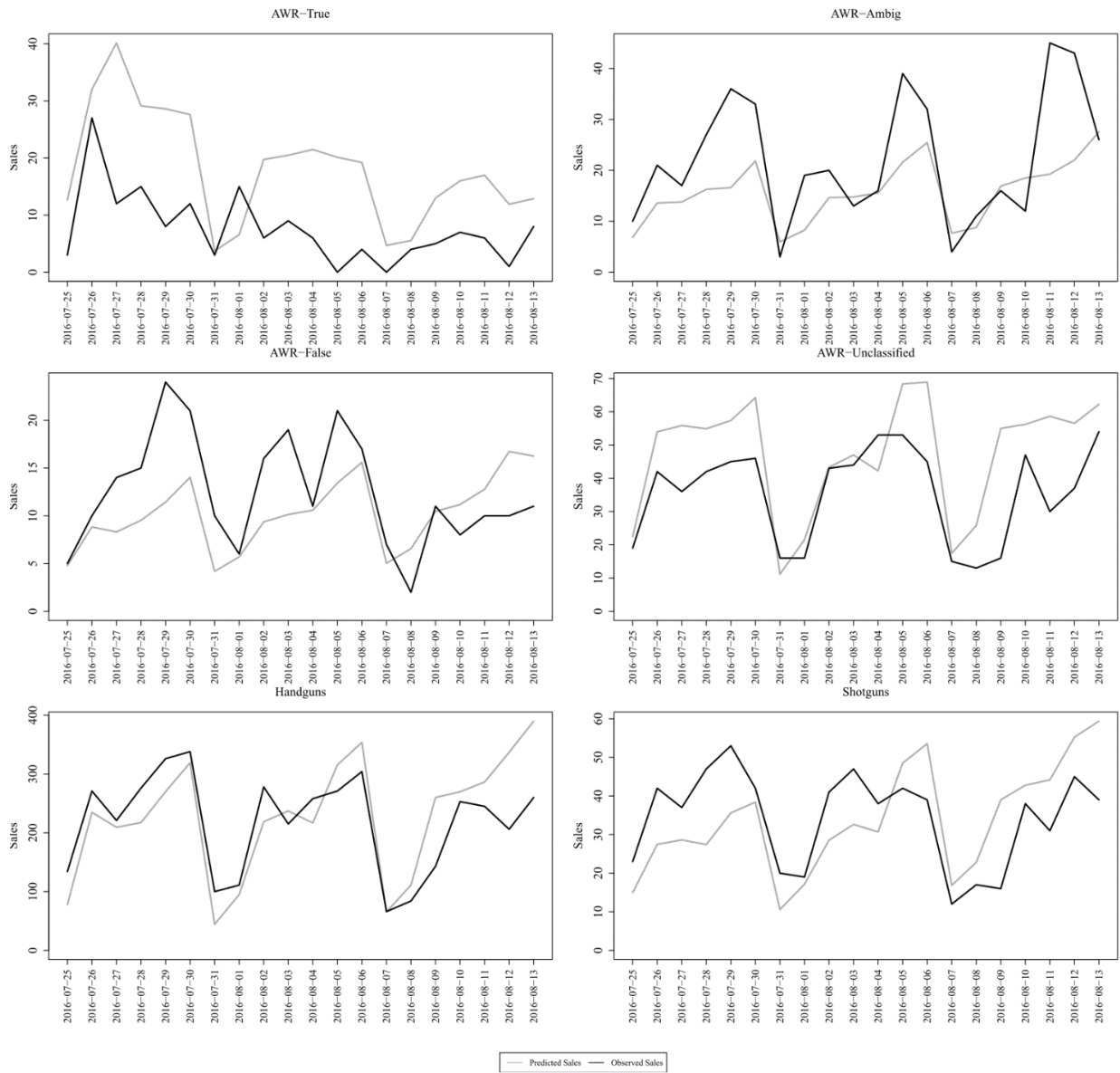

**Figure 9. Firearm sales in 2014-2017, by type and by week of year**

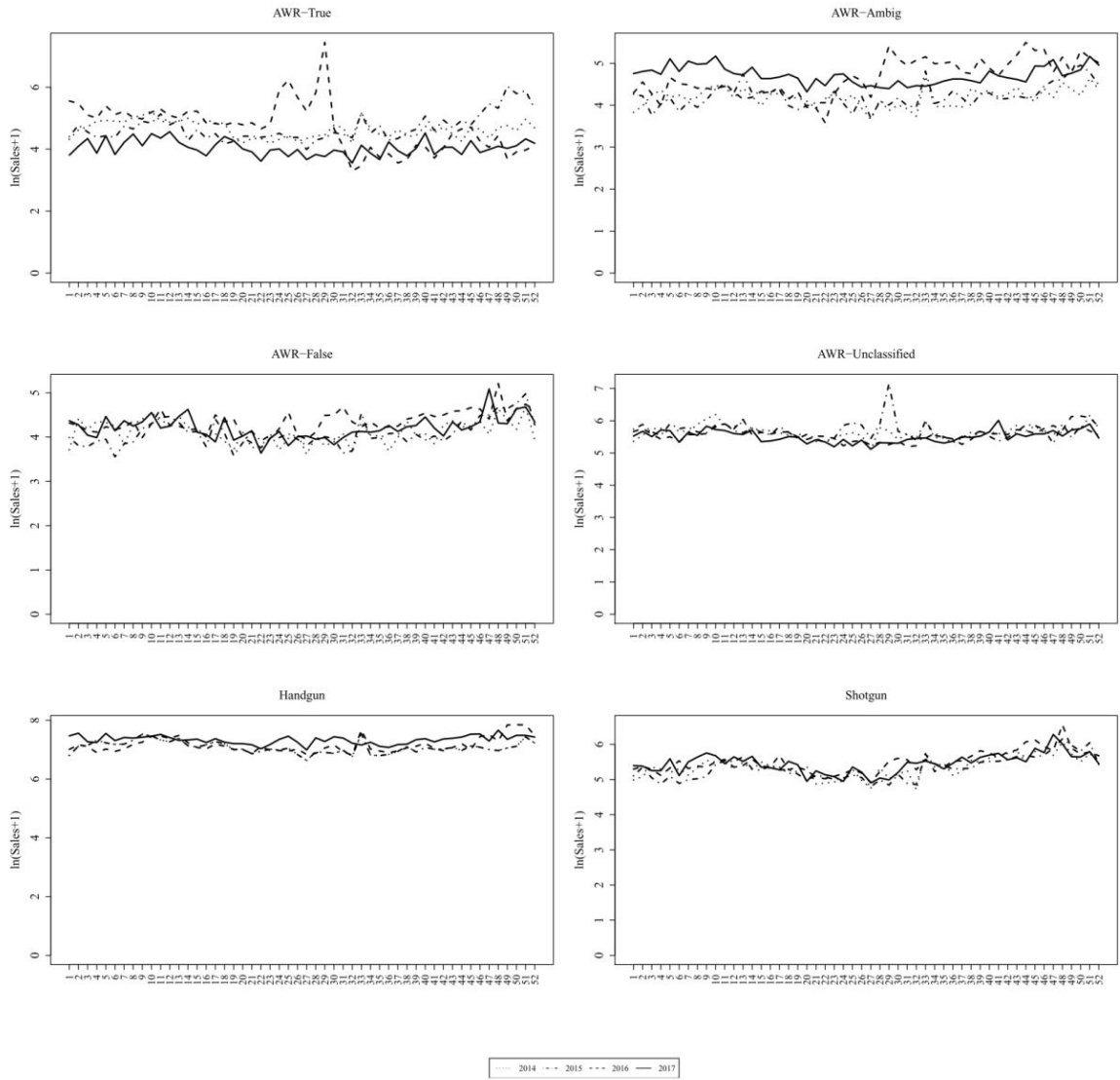

**Figure 10. Firearm Sales to Newly-observed Purchasers in 2014-2017, by Type and Week of Year**

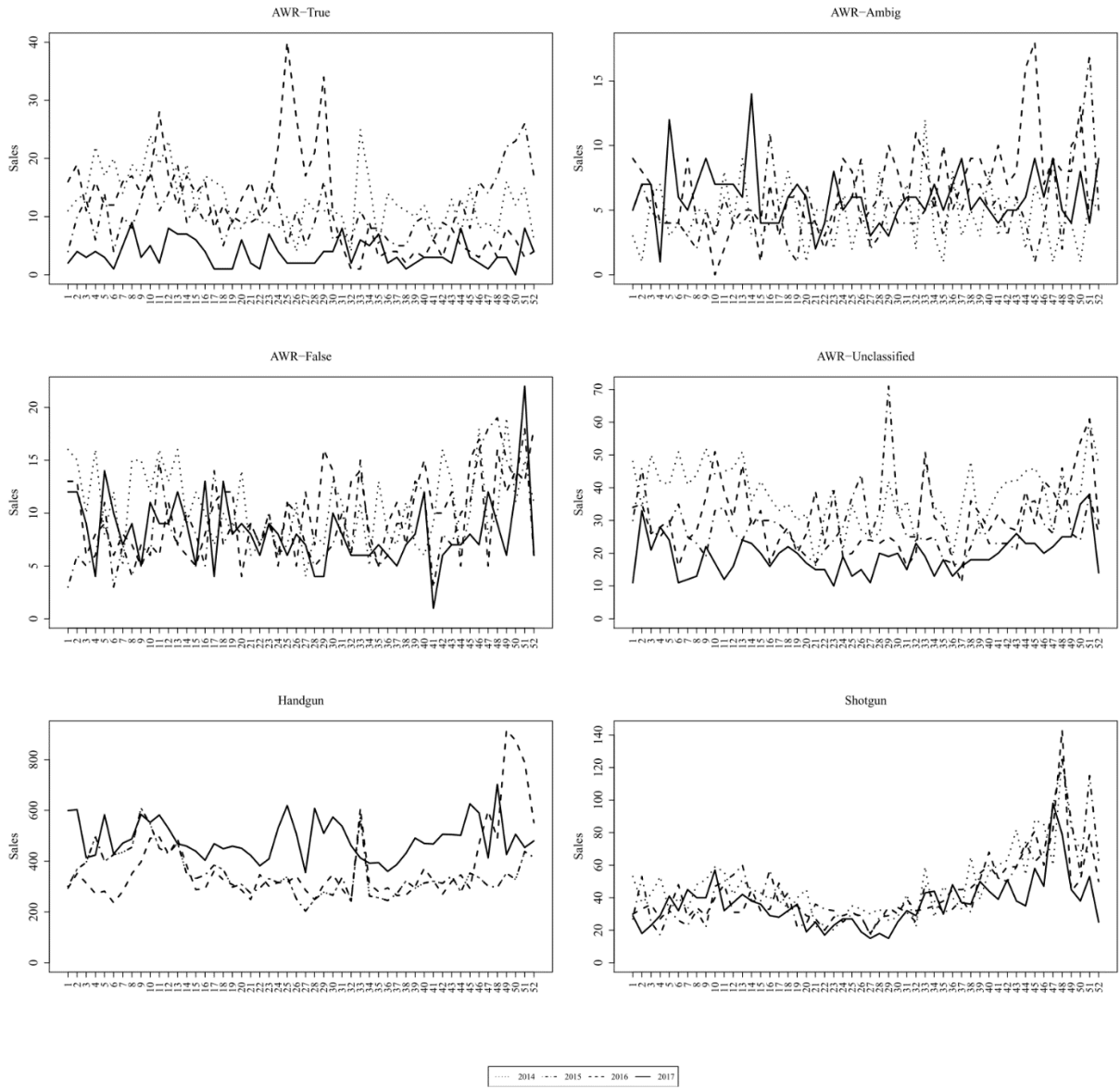

Appendix. Parameter Estimates.

| | Dependent variable: | | | | | |
|---|---|---|---|---|---|---|
| | AWR-True | AWR-Ambig | AWR-False | AWR-Unclassified | Handguns | Shotgun |
| Holiday | -0.55*** | -0.47*** | -0.45*** | -0.82*** | -1.12*** | -0.78*** |
| | (0.05) | (0.05) | (0.05) | (0.04) | (0.04) | (0.04) |
| Monday | -0.61*** | -0.69*** | -0.49*** | -0.85*** | -0.91*** | -0.61*** |
| | (0.10) | (0.09) | (0.09) | (0.09) | (0.09) | (0.09) |
| Tuesday | -0.21* | -0.24** | -0.1 | -0.31*** | -0.28*** | -0.22** |
| | (0.11) | (0.10) | (0.10) | (0.10) | (0.10) | (0.09) |
| Wednesday | -0.25** | -0.28** | -0.16 | -0.36*** | -0.36*** | -0.27*** |
| | (0.12) | (0.11) | (0.11) | (0.11) | (0.11) | (0.10) |
| Thursday | 0.02 | -0.07 | 0.01 | -0.16 | -0.28** | -0.16 |
| | (0.12) | (0.11) | (0.11) | (0.11) | (0.11) | (0.10) |
| Saturday | 0.26*** | 0.35*** | 0.27*** | 0.31*** | 0.47*** | 0.29*** |
| | (0.07) | (0.07) | (0.06) | (0.07) | (0.07) | (0.06) |
| Sunday | -0.84*** | -0.72*** | -0.56*** | -1.07*** | -1.11*** | -0.73*** |
| | (0.09) | (0.09) | (0.08) | (0.09) | (0.09) | (0.08) |
| Week 2 | -0.09 | 0.16 | 0.31* | 0.02 | 0.003 | 0.19 |
| | (0.17) | (0.16) | (0.16) | (0.15) | (0.15) | (0.15) |
| Week 3 | -0.18 | 0.1 | 0.32 | -0.05 | -0.21 | 0.2 |
| | (0.28) | (0.27) | (0.27) | (0.25) | (0.25) | (0.25) |
| Week 4 | -0.31 | 0.09 | 0.51 | -0.24 | -0.56 | 0.18 |
| | (0.39) | (0.38) | (0.38) | (0.35) | (0.36) | (0.35) |
| Week 5 | -0.29 | 0.17 | 0.55 | -0.38 | -0.75 | 0.29 |
| | (0.51) | (0.49) | (0.49) | (0.45) | (0.46) | (0.45) |
| Week 6 | -0.47 | 0.04 | 0.66 | -0.54 | -1.07* | 0.26 |
| | (0.63) | (0.60) | (0.60) | (0.56) | (0.56) | (0.55) |
| Week 7 | -0.41 | 0.27 | 0.94 | -0.42 | -0.97 | 0.55 |
| | (0.75) | (0.72) | (0.72) | (0.66) | (0.67) | (0.66) |
| Week 8 | -0.62 | 0.16 | 0.95 | -0.66 | -1.30* | 0.54 |
| | (0.86) | (0.83) | (0.83) | (0.76) | (0.78) | (0.76) |
| Week 9 | -0.62 | 0.24 | 1.16 | -0.74 | -1.47* | 0.57 |
| | (0.98) | (0.94) | (0.94) | (0.87) | (0.88) | (0.86) |
| Week 10 | -0.89 | 0.36 | 1.2 | -0.81 | -1.70* | 0.69 |
| | (1.10) | (1.06) | (1.05) | (0.97) | (0.99) | (0.97) |
| Week 11 | -0.98 | 0.23 | 1.37 | -0.99 | -2.00* | 0.68 |
| | (1.22) | (1.17) | (1.17) | (1.08) | (1.10) | (1.07) |
| Week 12 | -1.12 | 0.25 | 1.44 | -1.12 | -2.20* | 0.7 |
| | (1.33) | (1.28) | (1.28) | (1.18) | (1.20) | (1.17) |
| Week 13 | -1.14 | 0.32 | 1.5 | -1.16 | -2.35* | 0.73 |
| | (1.45) | (1.40) | (1.40) | (1.29) | (1.31) | (1.28) |
| Week 14 | -1.27 | 0.39 | 1.75 | -1.32 | -2.59* | 0.77 |
| | (1.57) | (1.51) | (1.51) | (1.39) | (1.41) | (1.38) |
| Week 15 | -1.36 | 0.37 | 1.75 | -1.46 | -2.72* | 0.91 |
| | (1.69) | (1.62) | (1.62) | (1.50) | (1.52) | (1.49) |
| Week 16 | -1.52 | 0.4 | 1.84 | -1.52 | -2.98* | 0.91 |
| | (1.81) | (1.74) | (1.74) | (1.60) | (1.63) | (1.59) |
| Week 17 | -1.58 | 0.36 | 1.93 | -1.58 | -3.01* | 1.04 |
| | (1.93) | (1.85) | (1.85) | (1.71) | (1.73) | (1.70) |
| Week 18 | -1.85 | 0.29 | 2.06 | -1.79 | -3.41* | 0.81 |
| | (2.04) | (1.97) | (1.96) | (1.81) | (1.84) | (1.80) |
| Week 19 | -1.94 | 0.33 | 2.14 | -2.01 | -3.73* | 0.85 |
| | (2.16) | (2.08) | (2.08) | (1.92) | (1.95) | (1.90) |
| Week 20 | -1.99 | 0.3 | 2.18 | -2.03 | -3.88* | 0.98 |
| | (2.28) | (2.19) | (2.19) | (2.02) | (2.05) | (2.01) |
| Week 21 | -2.15 | 0.32 | 2.38 | -2.2 | -4.13* | 0.87 |
| | (2.40) | (2.31) | (2.31) | (2.13) | (2.16) | (2.11) |
| Week 22 | -2.15 | 0.48 | 2.52 | -2.14 | -4.13* | 1.08 |
| | (2.52) | (2.42) | (2.42) | (2.23) | (2.27) | (2.22) |
| Week 23 | -2.24 | 0.55 | 2.71 | -2.28 | -4.41* | 1.2 |
| | (2.64) | (2.53) | (2.53) | (2.34) | (2.37) | (2.32) |
| Week 24 | -2.3 | 0.48 | 2.69 | -2.38 | -4.59* | 1.26 |
| | (2.76) | (2.65) | (2.65) | (2.44) | (2.48) | (2.42) |
| Week 25 | -2.39 | 0.51 | 2.88 | -2.55 | -4.84* | 1.22 |
| | (2.87) | (2.76) | (2.76) | (2.55) | (2.59) | (2.53) |
| Week 26 | -2.58 | 0.56 | 2.98 | -2.69 | -5.07* | 1.21 |
| | (2.99) | (2.88) | (2.87) | (2.65) | (2.69) | (2.63) |
| Week 27 | -2.76 | 0.31 | 2.93 | -2.91 | -5.41* | 1.21 |
| | (3.11) | (2.99) | (2.99) | (2.76) | (2.80) | (2.74) |

| | AWR-True | AWR-Ambig | AWR-False | AWR-Unclassified | Handguns | Shotgun |
|---|---|---|---|---|---|---|
| | | | | Dependent variable: | | |
| Week 28 | -2.76 | 0.67 | 3.2 | -2.7 | -5.29* | 1.49 |
| | (3.23) | (3.10) | (3.10) | (2.86) | (2.90) | (2.84) |
| Week 29 | -2.8 | 0.64 | 3.32 | -2.91 | -5.52* | 1.6 |
| | (3.35) | (3.22) | (3.21) | (2.97) | (3.01) | (2.94) |
| Week 30 | -2.96 | 0.68 | 3.45 | -3 | -5.71* | 1.58 |
| | (3.47) | (3.33) | (3.33) | (3.07) | (3.12) | (3.05) |
| Week 31 | -3.22 | 0.61 | 3.58 | -3.21 | -6.05* | 1.6 |
| | (3.58) | (3.45) | (3.44) | (3.18) | (3.23) | (3.15) |
| Week 32 | -3.11 | 0.77 | 3.77 | -3.1 | -6.06* | 1.88 |
| | (3.70) | (3.56) | (3.56) | (3.28) | (3.33) | (3.26) |
| Week 33 | -3.25 | 0.8 | 3.99 | -3.25 | -6.27* | 1.96 |
| | (3.82) | (3.67) | (3.67) | (3.39) | (3.44) | (3.36) |
| Week 34 | -3.43 | 0.54 | 3.89 | -3.53 | -6.72* | 1.8 |
| | (3.94) | (3.79) | (3.78) | (3.49) | (3.55) | (3.47) |
| Week 35 | -3.6 | 0.67 | 3.9 | -3.71 | -7.01* | 1.86 |
| | (4.06) | (3.90) | (3.90) | (3.60) | (3.65) | (3.57) |
| Week 36 | -3.62 | 0.76 | 4.2 | -3.73 | -7.04* | 1.86 |
| | (4.18) | (4.01) | (4.01) | (3.70) | (3.76) | (3.68) |
| Week 37 | -3.74 | 0.79 | 4.27 | -3.76 | -7.16* | 2.05 |
| | (4.29) | (4.13) | (4.12) | (3.81) | (3.86) | (3.78) |
| Week 38 | -3.77 | 0.87 | 4.48 | -3.78 | -7.23* | 2.19 |
| | (4.41) | (4.24) | (4.24) | (3.91) | (3.97) | (3.88) |
| Week 39 | -3.78 | 0.91 | 4.55 | -3.95 | -7.52* | 2.31 |
| | (4.53) | (4.36) | (4.35) | (4.02) | (4.08) | (3.99) |
| Week 40 | -4.01 | 0.85 | 4.74 | -4.07 | -7.69* | 2.29 |
| | (4.65) | (4.47) | (4.47) | (4.12) | (4.18) | (4.09) |
| Week 41 | -4.2 | 0.77 | 4.75 | -4.15 | -8.02* | 2.34 |
| | (4.77) | (4.58) | (4.58) | (4.23) | (4.29) | (4.20) |
| Week 42 | -4.18 | 0.8 | 4.93 | -4.3 | -8.19* | 2.41 |
| | (4.89) | (4.70) | (4.69) | (4.33) | (4.40) | (4.30) |
| Week 43 | -4.29 | 0.97 | 5.19 | -4.34 | -8.33* | 2.42 |
| | (5.01) | (4.81) | (4.81) | (4.44) | (4.50) | (4.41) |
| Week 44 | -4.23 | 1.19 | 5.37 | -4.32 | -8.36* | 2.57 |
| | (5.12) | (4.93) | (4.92) | (4.54) | (4.61) | (4.51) |
| Week 45 | -4.2 | 1.19 | 5.48 | -4.45 | -8.48* | 2.73 |
| | (5.24) | (5.04) | (5.03) | (4.65) | (4.72) | (4.61) |
| Week 46 | -4.61 | 1.02 | 5.4 | -4.76 | -8.98* | 2.53 |
| | (5.36) | (5.15) | (5.15) | (4.75) | (4.82) | (4.72) |
| Week 47 | -4.72 | 1.03 | 5.43 | -4.97 | -9.32* | 2.61 |
| | (5.48) | (5.27) | (5.26) | (4.86) | (4.93) | (4.82) |
| Week 48 | -4.68 | 1.18 | 5.84 | -4.83 | -9.24* | 2.77 |
| | (5.60) | (5.38) | (5.38) | (4.96) | (5.04) | (4.93) |
| Week 49 | -4.87 | 1.17 | 5.84 | -4.86 | -9.35* | 2.58 |
| | (5.72) | (5.49) | (5.49) | (5.06) | (5.14) | (5.03) |
| Week 50 | -4.75 | 1.31 | 6.09 | -4.82 | -9.40* | 2.7 |
| | (5.83) | (5.61) | (5.60) | (5.17) | (5.25) | (5.14) |
| Week 51 | -4.92 | 1.3 | 6.19 | -5.09 | -9.77* | 2.75 |
| | (5.95) | (5.72) | (5.72) | (5.28) | (5.36) | (5.24) |
| Week 52 | -5.17 | 1.12 | 6.08 | -5.4 | -10.26* | 2.61 |
| | (6.07) | (5.84) | (5.83) | (5.38) | (5.46) | (5.35) |
| Year 2007 | -5.32 | 1.21 | 6.12 | -5.52 | -10.27* | 2.69 |
| | (6.16) | (5.92) | (5.91) | (5.46) | (5.54) | (5.42) |
| Year 2008 | -10.56 | 2.38 | 12.27 | -10.99 | -20.49* | 5.43 |
| | (12.32) | (11.84) | (11.83) | (10.92) | (11.08) | (10.84) |
| Year 2009 | -15.88 | 3.52 | 18.37 | -16.53 | -30.78* | 8.1 |
| | (18.48) | (17.76) | (17.74) | (16.37) | (16.63) | (16.26) |
| Year 2010 | -21.25 | 4.64 | 24.39 | -22.09 | -41.10* | 10.74 |
| | (24.64) | (23.68) | (23.66) | (21.83) | (22.17) | (21.68) |
| Year 2011 | -26.62 | 5.87 | 30.65 | -27.71 | -51.58* | 13.47 |
| | (30.91) | (29.71) | (29.68) | (27.39) | (27.81) | (27.21) |
| Year 2012 | -31.88 | 7.11 | 36.86 | -33.16 | -61.88* | 16.16 |
| | (37.07) | (35.63) | (35.60) | (32.85) | (33.36) | (32.63) |
| Year 2013 | -37.25 | 8.18 | 42.96 | -38.7 | -72.26* | 18.81 |
| | (43.23) | (41.56) | (41.51) | (38.31) | (38.90) | (38.05) |
| Year 2014 | -42.69 | 9.24 | 48.91 | -44.19 | -82.52* | 21.49 |
| | (49.39) | (47.48) | (47.43) | (43.77) | (44.44) | (43.47) |
| Year 2015 | -48.09 | 10.32 | 54.99 | -49.67 | -92.77* | 24.19 |
| | (55.55) | (53.40) | (53.34) | (49.23) | (49.99) | (48.90) |
| Year 2016 | -53.41 | 11.43 | 61.1 | -55.16 | -103.13* | 26.85 |
| | (61.71) | (59.32) | (59.26) | (54.69) | (55.53) | (54.32) |
| Trend | 0.015 | -0.003 | -0.016 | 0.015 | 0.029* | -0.007 |
| | (0.02) | (0.02) | (0.02) | (0.02) | (0.02) | (0.02) |

| | Dependent variable: | | | | | |
|---|---|---|---|---|---|---|
| | AWR-True | AWR-Ambig | AWR-False | AWR-Unclassified | Handguns | Shotgun |
| Renewed License | 0.06** | 0.02 | -0.01 | 0.02 | 0.05** | 0.05** |
| | (0.03) | (0.02) | (0.02) | (0.02) | (0.02) | (0.02) |
| New license | 0.05** | 0.05** | 0.07*** | 0.11*** | 0.13*** | 0.07*** |
| | (0.02) | (0.02) | (0.02) | (0.02) | (0.02) | (0.02) |
| New licenses Lag 1 | -0.02 | 0.01 | 0.02 | -0.04* | -0.04 | -0.03 |
| | (0.02) | (0.02) | (0.02) | (0.02) | (0.02) | (0.02) |
| Renew licenses Lag 1 | -0.01 | -0.01 | -0.02 | 0.01 | 0.02 | 0.01 |
| | (0.03) | (0.02) | (0.02) | (0.02) | (0.02) | (0.02) |
| New licenses Lag 2 | 0.02 | -0.01 | 0.01 | -0.01 | -0.02 | -0.004 |
| | (0.02) | (0.02) | (0.02) | (0.02) | (0.02) | (0.02) |
| Renew licenses Lag 2 | -0.02 | 0.002 | -0.01 | 0.003 | 0.02 | 0.01 |
| | (0.02) | (0.02) | (0.02) | (0.02) | (0.02) | (0.02) |
| New licenses Lag 3 | 0.03 | -0.02 | -0.03 | -0.03 | -0.01 | -0.001 |
| | (0.02) | (0.02) | (0.02) | (0.02) | (0.02) | (0.02) |
| Renew licenses Lag 3 | -0.05** | -0.04 | -0.02 | -0.02 | -0.04* | -0.05** |
| | (0.02) | (0.02) | (0.02) | (0.02) | (0.02) | (0.02) |
| New licenses Lag 4 | 0.01 | 0.02 | 0.03 | 0.03 | 0.03 | 0.02 |
| | (0.02) | (0.02) | (0.02) | (0.02) | (0.02) | (0.02) |
| Renew licenses Lag 4 | -0.01 | 0.0005 | -0.02 | 0.01 | -0.01 | -0.01 |
| | (0.02) | (0.02) | (0.02) | (0.02) | (0.02) | (0.02) |
| New licenses Lag 5 | -0.02 | -0.02 | -0.03 | 0.01 | 0.02 | 0.03 |
| | (0.02) | (0.02) | (0.02) | (0.02) | (0.02) | (0.02) |
| Renew licenses Lag 5 | 0.03 | 0.01 | 0.04 | -0.02 | -0.03 | -0.02 |
| | (0.02) | (0.02) | (0.02) | (0.02) | (0.02) | (0.02) |
| New licenses Lag 6 | -0.03 | -0.02 | -0.01 | 0.01 | 0.03 | 0.01 |
| | (0.02) | (0.02) | (0.02) | (0.02) | (0.02) | (0.02) |
| Renew licenses Lag 6 | 0.02 | 0.02 | -0.01 | -0.01 | -0.01 | -0.003 |
| | (0.02) | (0.02) | (0.02) | (0.02) | (0.02) | (0.02) |
| New licenses Lag 7 | -0.004 | -0.003 | 0.01 | -0.02 | -0.01 | -0.01 |
| | (0.02) | (0.02) | (0.02) | (0.02) | (0.02) | (0.02) |
| Renew licenses Lag 7 | -0.03 | 0.01 | 0.01 | 0.01 | 0.003 | -0.003 |
| | (0.02) | (0.02) | (0.02) | (0.02) | (0.02) | (0.02) |
| New licenses Lag 8 | 0.01 | 0.002 | 0.02 | 0.02 | 0.02 | 0.01 |
| | (0.02) | (0.02) | (0.02) | (0.02) | (0.02) | (0.02) |
| Renew licenses Lag 8 | -0.001 | 0.004 | -0.02 | -0.002 | 0.001 | 0.01 |
| | (0.02) | (0.02) | (0.02) | (0.02) | (0.02) | (0.02) |
| New licenses Lag 9 | 0.01 | 0.03 | 0.005 | -0.01 | -0.01 | -0.002 |
| | (0.02) | (0.02) | (0.02) | (0.02) | (0.02) | (0.02) |
| Renew licenses Lag 9 | -0.01 | -0.04 | 0.01 | 0.01 | 0.02 | 0.0004 |
| | (0.02) | (0.02) | (0.02) | (0.02) | (0.02) | (0.02) |
| New licenses Lag 10 | -0.002 | 0.03 | -0.02 | 0.02 | 0.01 | 0.03 |
| | (0.02) | (0.02) | (0.02) | (0.02) | (0.02) | (0.02) |
| Renew licenses Lag 10 | -0.01 | -0.03 | 0.04 | -0.01 | 0.001 | -0.01 |
| | (0.02) | (0.02) | (0.02) | (0.02) | (0.02) | (0.02) |
| Constant | 0.13 | 0.42*** | 0.28** | 0.51*** | 0.66*** | 0.35** |
| | (0.14) | (0.14) | (0.14) | (0.16) | (0.21) | (0.15) |

| Dependent variable: | | | | | | | | | | | |
|---|---|---|---|---|---|---|---|---|---|---|---|
| AWR-True | | AWR-Ambig | | AWR-False | | AWR-Unclassified | | Handguns | | Shotgun | |
| AWR-True Lag 1 | 0.103*** (0.02) | AWR-Ambig Lag 1 | 0.041** (0.02) | AWR-False Lag 1 | 0.042*** (0.02) | AWR-Unclassified Lag 1 | 0.067*** (0.02) | Handgun Lag 1 | 0.025 (0.02) | Shotgun Lag 1 | 0.015 (0.02) |
| AWR-True Lag 2 | 0.036** (0.02) | AWR-Ambig Lag 2 | 0.018 (0.02) | AWR-False Lag 2 | 0.004 (0.02) | AWR-Unclassified Lag 2 | 0.023 (0.02) | Handgun Lag 2 | 0.001 (0.02) | Shotgun Lag 2 | 0.011 (0.02) |
| AWR-True Lag 3 | 0.070*** (0.02) | AWR-Ambig Lag 3 | 0.017 (0.02) | AWR-False Lag 3 | 0.016 (0.02) | AWR-Unclassified Lag 3 | 0.038** (0.02) | Handgun Lag 3 | 0.004 (0.02) | Shotgun Lag 3 | 0.001 (0.02) |
| AWR-True Lag 4 | 0.056*** (0.02) | AWR-Ambig Lag 4 | 0.013 (0.02) | AWR-False Lag 4 | -0.007 (0.02) | AWR-Unclassified Lag 4 | 0.005 (0.02) | Handgun Lag 4 | -0.014 (0.02) | Shotgun Lag 4 | -0.007 (0.02) |
| AWR-True Lag 5 | 0.008 (0.02) | AWR-Ambig Lag 5 | -0.0001 (0.02) | AWR-False Lag 5 | 0.009 (0.02) | AWR-Unclassified Lag 5 | 0.014 (0.02) | Handgun Lag 5 | -0.0004 (0.02) | Shotgun Lag 5 | 0.013 (0.02) |
| AWR-True Lag 6 | 0.058*** (0.02) | AWR-Ambig Lag 6 | 0.01 (0.02) | AWR-False Lag 6 | 0.040** (0.02) | AWR-Unclassified Lag 6 | 0.038*** (0.02) | Handgun Lag 6 | 0.014 (0.02) | Shotgun Lag 6 | 0.037** (0.02) |
| AWR-True Lag 7 | 0.110*** (0.02) | AWR-Ambig Lag 7 | 0.044*** (0.02) | AWR-False Lag 7 | 0.062*** (0.02) | AWR-Unclassified Lag 7 | 0.057*** (0.02) | Handgun Lag 7 | 0.074*** (0.02) | Shotgun Lag 7 | 0.079*** (0.02) |
| AWR-True Lag 8 | 0.019 (0.02) | AWR-Ambig Lag 8 | 0.051*** (0.02) | AWR-False Lag 8 | 0.027* (0.02) | AWR-Unclassified Lag 8 | 0.030* (0.02) | Handgun Lag 8 | 0.029* (0.02) | Shotgun Lag 8 | 0.029* (0.02) |
| AWR-True Lag 9 | 0.005 (0.02) | AWR-Ambig Lag 9 | 0.023 (0.02) | AWR-False Lag 9 | 0.016 (0.02) | AWR-Unclassified Lag 9 | 0.036** (0.02) | Handgun Lag 9 | 0.047*** (0.02) | Shotgun Lag 9 | 0.043*** (0.02) |
| AWR-True Lag 10 | 0.014 (0.02) | AWR-Ambig Lag 10 | 0.013 (0.02) | AWR-False Lag 10 | 0.003 (0.02) | AWR-Unclassified Lag 10 | 0.007 (0.02) | Handgun Lag 10 | 0.024 (0.02) | Shotgun Lag 10 | -0.009 (0.02) |
| AWR-True Lag 11 | 0.017 (0.02) | AWR-Ambig Lag 11 | 0.012 (0.02) | AWR-False Lag 11 | -0.002 (0.02) | AWR-Unclassified Lag 11 | 0.02 (0.02) | Handgun Lag 11 | 0.008 (0.01) | Shotgun Lag 11 | 0.034** (0.02) |
| AWR-True Lag 12 | -0.004 (0.02) | AWR-Ambig Lag 12 | 0.049*** (0.02) | AWR-False Lag 12 | 0.029* (0.02) | AWR-Unclassified Lag 12 | 0.037** (0.02) | Handgun Lag 12 | 0.031** (0.01) | Shotgun Lag 12 | 0.034** (0.02) |
| AWR-True Lag 13 | 0.003 (0.02) | AWR-Ambig Lag 13 | 0.043*** (0.02) | AWR-False Lag 13 | 0.011 (0.02) | AWR-Unclassified Lag 13 | 0.030** (0.02) | Handgun Lag 13 | 0.047*** (0.01) | Shotgun Lag 13 | 0.018 (0.02) |
| AWR-True Lag 14 | 0.067*** (0.02) | AWR-Ambig Lag 14 | 0.018 (0.02) | AWR-False Lag 14 | 0.024 (0.02) | AWR-Unclassified Lag 14 | 0.025 (0.02) | Handgun Lag 14 | 0.026* (0.01) | Shotgun Lag 14 | 0.033** (0.02) |
| AWR-True Lag 15 | 0.031* (0.02) | AWR-Ambig Lag 15 | 0.034** (0.02) | AWR-False Lag 15 | 0.017 (0.02) | AWR-Unclassified Lag 15 | 0.011 (0.02) | Handgun Lag 15 | 0.027* (0.01) | Shotgun Lag 15 | 0.041*** (0.02) |
| AWR-True Lag 16 | -0.021 (0.02) | AWR-Ambig Lag 16 | 0.017 (0.02) | AWR-False Lag 16 | 0.013 (0.02) | AWR-Unclassified Lag 16 | 0.023 (0.02) | Handgun Lag 16 | 0.021 (0.02) | Shotgun Lag 16 | 0.018 (0.02) |
| AWR-True Lag 17 | 0.014 (0.02) | AWR-Ambig Lag 17 | 0.025 (0.02) | AWR-False Lag 17 | 0.003 (0.02) | AWR-Unclassified Lag 17 | 0.018 (0.02) | Handgun Lag 17 | 0.02 (0.02) | Shotgun Lag 17 | 0.039** (0.02) |
| AWR-True Lag 18 | 0.011 (0.02) | AWR-Ambig Lag 18 | 0.022 (0.02) | AWR-False Lag 18 | 0.045*** (0.02) | AWR-Unclassified Lag 18 | 0.036** (0.02) | Handgun Lag 18 | 0.044*** (0.02) | Shotgun Lag 18 | 0.02 |
| AWR-True Lag 19 | 0.011 (0.02) | AWR-Ambig Lag 19 | 0.012 (0.02) | AWR-False Lag 19 | 0.012 (0.02) | AWR-Unclassified Lag 19 | 0.026* (0.02) | Handgun Lag 19 | 0.022 (0.02) | Shotgun Lag 19 | 0.030** (0.02) |
| AWR-True Lag 20 | 0.023 (0.02) | AWR-Ambig Lag 20 | 0.013 (0.02) | AWR-False Lag 20 | 0.022 (0.02) | AWR-Unclassified Lag 20 | 0.037** (0.02) | Handgun Lag 20 | 0.023 (0.02) | Shotgun Lag 20 | 0.026* (0.02) |
| AWR-True Lag 21 | 0.035* (0.02) | AWR-Ambig Lag 21 | 0.045*** (0.02) | AWR-False Lag 21 | 0.025 (0.02) | AWR-Unclassified Lag 21 | 0.040*** (0.02) | Handgun Lag 21 | 0.074*** (0.02) | Shotgun Lag 21 | 0.053*** (0.02) |
| AWR-True Lag 22 | 0.045*** (0.02) | AWR-Ambig Lag 22 | 0.011 (0.02) | AWR-False Lag 22 | 0.022 (0.02) | AWR-Unclassified Lag 22 | 0.023 (0.02) | Handgun Lag 22 | 0.030** (0.02) | Shotgun Lag 22 | 0.028* (0.02) |
| AWR-True Lag 23 | -0.029* (0.02) | AWR-Ambig Lag 23 | 0.007 (0.02) | AWR-False Lag 23 | -0.0004 (0.02) | AWR-Unclassified Lag 23 | 0.002 (0.02) | Handgun Lag 23 | 0.015 (0.01) | Shotgun Lag 23 | 0.02 (0.02) |
| AWR-True Lag 24 | 0.031* (0.02) | AWR-Ambig Lag 24 | 0.008 (0.02) | AWR-False Lag 24 | 0.013 (0.02) | AWR-Unclassified Lag 24 | 0.001 (0.02) | Handgun Lag 24 | 0.030** (0.02) | Shotgun Lag 24 | 0.007 (0.02) |
| AWR-True Lag 25 | -0.005 (0.05) | AWR-Ambig Lag 25 | -0.006 (0.02) | AWR-False Lag 25 | -0.012 (0.02) | AWR-Unclassified Lag 25 | 0.012 (0.02) | Handgun Lag 25 | -0.013 (0.02) | Shotgun Lag 25 | -0.01 (0.02) |
| AWR-True Lag 26 | -0.022 (0.02) | AWR-Ambig Lag 26 | 0.013 (0.02) | AWR-False Lag 26 | -0.007 (0.02) | AWR-Unclassified Lag 26 | 0.006 (0.02) | Handgun Lag 26 | 0.016 (0.01) | Shotgun Lag 26 | 0.031** -0.015 |
| AWR-True Lag 27 | 0.026 (0.02) | AWR-Ambig Lag 27 | 0.011 (0.02) | AWR-False Lag 27 | 0.007 (0.02) | AWR-Unclassified Lag 27 | -0.001 (0.02) | Handgun Lag 27 | 0.004 (0.02) | Shotgun Lag 27 | 0.012 (0.02) |
| AWR-True Lag 28 | 0.064*** (0.02) | AWR-Ambig Lag 28 | 0.003 (0.02) | AWR-False Lag 28 | 0.003 (0.02) | AWR-Unclassified Lag 28 | 0.055*** (0.02) | Handgun Lag 28 | 0.037*** -0.014 | Shotgun Lag 28 | 0.051*** (0.02) |